\documentclass[twocolumn,showpacs,preprintnumbers,amsmath,amssymb]{revtex4}

\usepackage{graphicx}
\usepackage{dcolumn}
\usepackage{bm}

\begin{document}

\title{Nonlinear parametric amplification in a tri-port nanoelectromechanical device}

\author{E. Collin} 
\email{eddy.collin@grenoble.cnrs.fr}
\author{T. Moutonet}
\author{J.-S. Heron}
\author{O. Bourgeois}
\author{Yu. M. Bunkov}
\author{H. Godfrin}

\affiliation{%
Institut N\'eel
\\
CNRS et Universit\'e Joseph Fourier, \\
BP 166, 38042 Grenoble Cedex 9, France \\
}%

\date{\today}

\begin{abstract}
We report on measurements performed at low temperatures on a nanoelectromechanical system (NEMS) under (capacitive) parametric pumping. The excitations and detection schemes are purely electrical, and enable in the present experiment the straightforward measurement of forces down to about a femtonewton, for displacements of an Angstr\"om, using standard room temperature electronics.
 We demonstrate that a small (linear) force applied on the device can be amplified up to more than a 100 times, while the system is {\it truly moving}. We explore the dynamics up to about 50$~$nm deflections for cantilevers about 200$~$nm thick by 3$~$$\mu$m long oscillating at a frequency of 7$~$MHz. We present a generic modeling of nonlinear parametric amplification, and give analytic theoretical solutions enabling the fit of experimental results. We finally discuss the practical limits of the technique, with a particular application: the measurement of {\it anelastic damping} in the metallic coating of the device with an exceptional resolution of about 0.5$~$\%. 

\end{abstract}

\maketitle

\section{INTRODUCTION}

Condensed matter physicists have always been concerned with the measurement of small forces (arising from the interactions between  constitutive elements of the systems they study), or small masses (detecting the presence of these actual constitutive elements). With the amazing advance in fabrication technologies, micromechanical and nanomechanical probes have become available extending the detection limits down to extremely weak forces, and small quantities of matter. Many applications have emerged, for instance sensitive accelerometers \cite{accel}, the detection of a single electron spin \cite{rugarnature}, a 90 nm scale resolution Magnetic Resonance Force Microscope (MRFM) \cite{rugarnucl}, the characterization of nanostructures with the Atomic Force Microscope (AFM) \cite{AFM}, Electric Force Microscopy (EFM) \cite{EFM}, the detection of single electronic charges \cite{roukescleleand}, the precise measurement of the Casimir force \cite{mohideen} and finally attogram mass sensing \cite{mass1,mass2} with potential applications in chemical and biological sensing \cite{chemical,bio}.

Ultimately, the force sensitivity of a nanomechanical probe detected through a linear amplifier will be limited by the {\it standard quantum limit} sought in mechanical quantum ground state cooling \cite{schwab}.
To date, the world record on force sensing using a nanomechanical object is about an attonewton, with a laser readout \cite{attonewton} or a shot-noise limited microwave interferometer \cite{lehnert}, both operated at cryogenic temperatures; using a crystal of trapped ions this limit was recently pushed down to the incredibly low value of about 400 yoctonewton \cite{NIST}.

The operating principle of most of the devices is based on the measurement of a mechanical resonance mode which frequency $f_0$, quality factor $Q$ or oscillation phase $\varphi$ depends on the physical property one wishes to extract (force, or additional mass). The force sensitivity of such a setup which is limited by the detection noise floor ultimately relies on the quality factor $Q$ of the mechanical resonance mode under use, defined as the ratio of the energy of the motion to the energy lost per oscillation cycle. Three strategies can be considered to improve the $Q$:
\begin{itemize}
\item design the geometry of the device and chose the constituents to minimize the damping (ultimately clamping losses \cite{clamp} and thermoelastic damping \cite{thermoel}) with respect to the energy stored in the resonance mode \cite{jeevakN,highQ}.
\item making use of a fast electronic feedback circuit, the so-called "Q-Control" scheme \cite{tamayo}.
\item making use of a parametric drive scheme, as first proposed for micromechanical devices in Ref. \cite{rugarPRL}.
\end{itemize}
These choices are not incompatible with each other: parametric schemes can also be used on {\it already very good} oscillators in order to improve them even further. On the other hand, it is usually agreed that because of transient signals, "Q-Control" reduces the speed at which one can change the working point of the oscillator \cite{qcontrol}. It also increases the background noise in the device \cite{tamayo}. 

Parametric drive schemes have thus gained a lot of attention in recent years. While the original Ref. \cite{rugarPRL} was dealing with parametric amplification only, the interest of the physicists has evolved towards parametric oscillation \cite{jeevakvanderpohl,fiveparam,nanowires}, incorporating also electronic feedback schemes to it \cite{stmparam}.\\
Parametric oscillation is the mechanism underlying the child's swing motion \cite{swing} and surface wave patterns \cite{water}. Before being implemented in micro and nanomechanical devices, it has been applied with great success in classical electronic circuits \cite{elec}, superconducting circuits \cite{josephson} and optical circuits \cite{opo}. In the latter case, the light field possesses remarkable quantum properties, and the technique is still one of the most widely used to generate squeezed coherent states and entangled states of photons \cite{squeezedlight}. 

Parametric oscillation (or parametric resonance, autoparametric drive) occurs when some parameter of a mechanical device with a mode resonance at $\omega_0=2 \pi f_0$, usually the spring constant $k$, is modulated at a frequency around $2 \omega_0/n$ ($n$ integer), with an amplitude $\Delta k$ exceeding a critical value. The system spontaneously oscillates at $\omega_0$, without any additional external force \cite{book,book2}. The great outcome of this particular drive scheme is that the resonance line obtained is {\it much narrower} than in a standard (linear) excitation, giving rise to greater effective $Q$ factors as already stated  \cite{jeevakvanderpohl,stmparam,pumphighQ}.

If the spring constant modulation amplitude (or "parametric pumping") is smaller than the critical value, the mechanical device is in the parametric amplification regime \cite{book,book2}. As experimentally demonstrated in Ref. \cite{rugarPRL}, the mechanical response to an external AC force (oscillating at $\omega_0$) can be either amplified (with a gain $> 1$), or squeezed (gain $< 1$), depending on {\it the phase} of the spring constant modulation with respect to the force. In this regime, the linewidth of the resonance line can also be substantially reduced, depending on the strength of the pumping. Squeezing can be used to reduce the thermomechanical noise of the oscillator in one of its quadratures \cite{rugarPRL}.

In practice, experiments are performed with the simple choice $n=1$, with the pump frequency $\omega_p$ around $2\omega_0$ and the excitation force frequency $\omega$ around $\omega_0$, both within the linewidth of the resonance $\Delta \omega$.
This is called {\it degenerate} parametric drive \cite{book2}.
 In order to modulate the spring constant $k$, one needs a nonlinear force or a nonlinear elastic property that can be tuned with an external drive.

The first realization of parametric amplification (and squeezing) in a micromechanical device was reported in Ref. \cite{rugarPRL}. The device was a 500$~\mu$m by 10$~\mu$m by a few micron thick silicon cantilever (resonating at 33$~$kHz), and gains up to 25 were reported, for displacements up to about 30 nm. The excitation was piezoelectric, the pumping was due to the nonlinear force generated by an additional capacitive drive, and the readout was optical (interferometric setup). \\
Macroscopic mechanical amplification was addressed in Ref. \cite{macro} with a steel cantilever 190$~$mm $\times$ 19$~$mm $\times$ 0.5$~$mm resonating at the very low frequency of 11.5$~$Hz. Squeezing was reported, and amplification gains of 1.6 were achieved. \\
Parametric amplification was obtained in a GaAs piezoresistive cantilever developed for Atomic Force Microscopy \cite{gaasstress}.
The structure was 230$~\mu$m long, 23$~\mu$m wide and 0.95$~\mu$m thick with a resonant frequency of 11.3$~$kHz. 
The cantilever was driven with a  piezoelectric PZT actuator while the parametric effect was realized through in-built stress which curves the beam and generates a second order nonlinear spring constant. The piezoresistive effect of the structure was used to measure the deflection.
In this experiment, amplifier gains up to 10 were measured, and the squeezing was demonstrated. Displacements of the order of 60$~$nm were realized without reducing the gain. \\
Higher frequencies (around 485$~$kHz) were obtained in torsional oscillators, with silicon bars of typically 200$~$nm by 200$~$nm by 1$~\mu$m, and paddles of about $2 \times 2 ~\mu$m$^2$ by 200$~$nm \cite{torsionalparpia}. The experimental scheme used a capacitive drive and pump, with an optical detection. The maximal reported gain was about 10. Squeezing was demonstrated.\\
The Megahertz range (0.9 MHz) was attained with a silicon disk oscillator \cite{parpiadisk} (20$~\mu$m by 250$~$nm on a 1$~\mu$m SiO$_2$ pillar of about 7$~\mu$m diameter). The device was piezoelectrically driven, optically pumped (through the generation of thermal gradients) and detected. Both amplification and squeezing were realized, with a highest reported gain of about 33. \\
The highest parametric oscillator frequency ever achieved was about 130 MHz in Ref. \cite{roukeshighfreq}, using a composite SiC+Al 2$~\mu$m long by 100$~$nm thick and wide beam. The experiments used the magnetomotive scheme for the drive and detection, and were performed at 10$~$K with 8$~$T magnetic fields. The parametric pumping was obtained through the modulation of the tensile stress in the beam, which was achieved with the magnetomotive distortion of "clamping rods" supporting the beam. The maximal gains reported were about 3. \\
An efficient parametric amplification scheme was demonstrated in coupling dispersively a nanomechanical beam to a Cooper Pair Box (CPB) quantum bit \cite{schwabqubit}. Both amplification and squeezing can be achieved, and maximal gains around 30 were reported. The particularity of the technique, which demonstrate very high coupling between the nanomechanical beam and the Cooper Pair Box, is that while the effect is purely classical, the quantum bit is {\it truly} quantum, and should enable in future works the quantum steering of the beam.  \\
Finally, the largest gains were attained in Ref. \cite{highgain}, for a 10$~$MHz, 8$~\mu$m long by 500$~$nm and 200$~$nm doped epitaxial GaAs beam. The actuation was piezoelectric, together with the parametric pumping which was obtained by modulating the longitudinal stress in the beam. The detection scheme was optical. 
The best gain reported was of a 1000, but for displacements of 14$~$pm. At displacements of a 100$~$pm, it was already below the value of 100.

Clearly in most experiments, the devices have extremely small displacements (as compared to their size), and often the detection requires very sensitive interferometric techniques. In this regime, experimentalists rely on the {\it linear} theory of parametric amplification \cite{book,book2,book3}: all high-order terms in the dynamics equation (like the well-known Duffing force $F_{non-lin.}=- k_3 x^3$) are taken to be zero and the only nonlinear term is the (first order) parametric modulation itself. Theoretical works based on linear parametric amplification have also been released for Electric Force Microscopy \cite{chevrier}, or mass sensing \cite{cleland}. 
On the other hand, work has been done on both the theoretical and experimental sides exploring the {\it nonlinear} regime of parametric oscillations \cite{nlinpaper0,nlinpaper1,nlinpaper2,nlinpaper3,nlinpaper4}. \\
It is of great importance to realize and study the large deflection amplification range. 
In the first place, in order to provide mechanical parametric amplifiers which are easy to handle, i.e. which {\it move enough} to be detected with simple electric setups: fully micro or nanoelectromechanical systems (MEMS or NEMS).
Secondly, in order to understand the limits on the gain arising from large deflections. In the linear theory, the gain is infinite at the threshold between amplification and oscillation. In practice, actual nonlinear effects will limit the gain \cite{book2,highgain}. The practical gain limitations where considered only in Ref. \cite{roukeshighfreq}, where the importance of the thermal stability of the $Q$ factor was addressed. 
A complete theory of nonlinear amplification is thus needed both for design purposes, and for a detailed physical understanding. 

In the present article we report on experiments carried on nanoelectromechanical devices (200$~$nm thick, 3$~\mu$m long cantilevers) presenting large gains and large displacements. Gains slightly above a 100 have been measured, and displacements ranging from 0.15 nanometer to about 50$~$nm have been explored. The highest gain could be achieved with a structure moving about a couple of \% of its thickness (peak-to-peak), which is exceptional. The experiment is performed at helium temperatures with an extremely simple setup involving three coaxial lines connected to commercial room temperature equipment. The parametric pumping is obtained via a straightforward capacitive coupling, and the (linear) "test force" is realized with the magnetomotive scheme. The magnetomotive induced voltage enables to detect the displacement of the structure as well. We give analytic mathematical tools to descibe the experimental results. Finally, we discuss the practical limits of parametric amplification, and present an interesting new application: the characterization of anelastic properties of the metallic coatings of NEMS through extremely precise measurements of the $Q$ factor of the resonance (about $\pm$0.5$~$\% resolution).

\section{EXPERIMENTAL SETUP} 
\label{setups}

\begin{figure}[t!]
\includegraphics[height=8.5 cm]{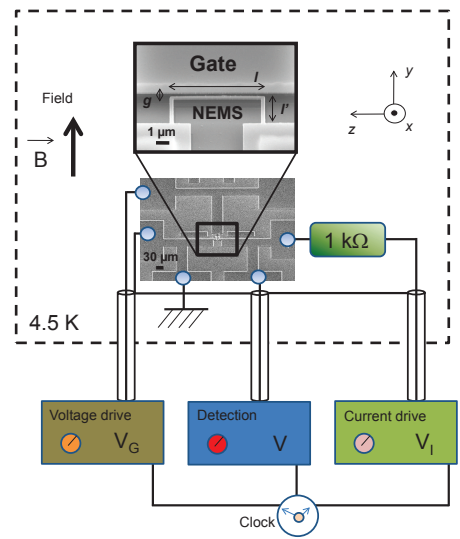}
\caption{\label{setup} (Color online) 
Schematic drawing of the experimental setup, with Scanning Electron Microscope (SEM) image of the sample. Only the chip and the drive resistor are maintained at cryogenic temperatures (dashed box); three coaxial cables are required to connect the tri-port device to conventional room temperature electronics. }
\end{figure}

The experimental setup is depicted in Fig. \ref{setup}, and has been presented in Ref. \cite{qfs}. 
It consists in a goal-post shaped NEMS device (two "feet" of length 3.1$~\mu$m plus a "paddle" of 7$~\mu$m, 280$~$nm wide) fabricated on a thick SOI substrate. The oxide layer was 1$~\mu$m thick for a top silicon thickness of 150$~$nm. The structure is patterned by means of Reactive Ion Etching (SF$_6$ plus O$_2$ plasma) through an Al sacrificial mask obtained by e-beam lithography. The beams are released after HF chemical vapor etching. About 30$~$nm of aluminum has been deposited on top of the structure to create electrical contacts. This metallic layer is rather soft, leaving the structure unstressed and adding only little elasticity.

The mobile part is composed of the two cantilever feet linked by the paddle. Due to symmetry, in its first resonant mode (out-of-plane flexural mode) the device behaves as a simple cantilever loaded by half the paddle at its extremity. This geometry has been extensively studied in our group, from MEMS to NEMS scales with various metallic coatings \cite{qfs,JLTP_VIW,qfs2009,coatings,nonlinPRB}.\\
The sample is placed in a helium cryostat, in a small cell in which cryogenic vacuum is maintained (typically less than 10$^{-6}~$mbar). Experiments are performed at 4.5$\pm 0.25~$K, taking into account self-heating of the structure (see below). 

This tri-port device is a very simple model system for the study of parametric amplification. The probe force to be amplified is realized by the magnetomotive scheme, which is by construction a {\it linear} excitation (independent of the displacement $x$): the sample is placed in a magnetic field $B$ (smaller than 1 T in our experiments) parallel to the cantilever feet, while a current $I(t)=I_0 \cos (\omega t)$ is fed through it with a cold bias resistor (voltage $V_I(t)$ across 1$~$k$\Omega$). The NEMS metallic layer has been measured to be 110$~\Omega$ at 4.2$~$K. The resulting force is perpendicular to the chip surface and writes $F(t)=I(t) l B$, with $l$ the length of the paddle. \\
The detection of the motion is carried out through the measurement of the induced voltage $V(t)$ appearing at the structure's ends while it moves and cuts the field lines. This output voltage can be shown to be {\it almost linear}, reducing to the expression $V(t)=l B \dot{x}(t)$, $x$ being the coordinate of the paddle position \cite{JLTP_VIW,note}.\\
The nonlinear force enabling the parametric pumping is realized with a gate electrode positioned 100$~$nm next to the paddle \cite{qfs}. The total gate capacitance $C_t$ to ground can be written:
\begin{displaymath}
C_t = C_0 +  \int_{NEMS} \delta C,
\end{displaymath}
with $C_0$ the parasitic capacitance due to the leads and connecting pads, and $\int_{NEMS} \delta C$ is the NEMS metallic layer contribution. Since the capacitance falls to zero extremely quickly with the distance to the gate electrode, in the latter integral we can keep only the paddle contribution.  
Parameterizing the NEMS capacitance with the global position of the paddle in the $\vec{x}$ and $\vec{y}$ directions (see Fig. \ref{setup}) leads to:
\begin{displaymath}
 \int_{NEMS} \delta C =C(x,y).
\end{displaymath}
The potential energy originating from the voltage bias $V_G$ on the gate electrode writes then:
\begin{displaymath}
E_C = \frac{1}{2} C(x,y) V_G^2.
\end{displaymath}
The resulting force (on the mobile part) is thus:
\begin{displaymath}
\vec{F}_C = +\frac{1}{2} \left( \frac{\partial C(x,y)}{\partial x} \vec{x} + \frac{\partial C(x,y)}{\partial y} \vec{y} \right) V_G^2.
\end{displaymath}
Keeping in mind that the displacements $x,y$ are small (in practice smaller than half the electrode's gap, i.e. 50$~$nm), we proceed with the following Taylor series expansions:
\begin{eqnarray*}
\frac{\partial C(x,y)}{\partial x} & = & \frac{\partial C(0,0)}{\partial x}+\frac{\partial^2 C(0,0)}{\partial x^2} x \\
&+ & \frac{1}{2} \frac{\partial^3 C(0,0)}{\partial x^3} x^2 + \frac{\partial^2 C(0,0)}{\partial x \partial y} y \\
&+ & \frac{\partial^3 C(0,0)}{\partial x^2 \partial y} x y + \frac{1}{6} \frac{\partial^4 C(0,0)}{\partial x^4} x^3 ,  \\
\frac{\partial C(x,y)}{\partial y} & = & \frac{\partial C(0,0)}{\partial y}+\frac{\partial^2 C(0,0)}{\partial x \partial y} x \\
&+ & \frac{1}{2} \frac{\partial^3 C(0,0)}{\partial x^2 \partial y} x^2 + \frac{\partial^2 C(0,0)}{\partial y^2} y \\
&+ & \frac{\partial^3 C(0,0)}{\partial x \partial y^2} x y  + \frac{1}{6} \frac{\partial^4 C(0,0)}{\partial x^3 \partial y} x^3,
\end{eqnarray*}
where we kept only orders smaller than 3. The $y$ displacement is at lowest order a $x$ second order \cite{JLTP_VIW,note}; we estimate $y \approx -  3/5 \, x^2/l'$ ($l'$ length of feet).

The $\vec{y}$ component of $\vec{F}_C$ (denoted $F^y_C$) is an axial force load acting on the feet of the structure. It influences their spring constants $k_{foot}$, and we have at first order \cite{JLTP_VIW,coatings}:  
\begin{displaymath}
k_{foot} = \frac{k_0}{2} \left( 1 - \sigma \frac{F^y_C/2 \, l'^2}{E \, I_y} \right) 
\end{displaymath}
with $k_0/2$ the unaxially-loaded spring constant of each foot, $E$ the corresponding Young modulus and $I_y$ its second moment of area.
$\sigma$ is a small number that is estimated for our geometry to be about $+0.095$. The end mass load due to the paddle is taken into account in this writing \cite{coatings}. \\
The $\vec{x}$ component $F^x_C$ of the force generates directly a modulation of the NEMS restoring force. Combining the axial load $F^y_C$ with the latter, we finally obtain the {\it effective} out-of-plane gate contribution:
\begin{eqnarray}
F_C^{ef\!f.} & = & +\frac{1}{2} \frac{\partial C(0,0)}{\partial x}  V_G^2 \nonumber \\
&  &\!\!\!\!\!\!\!\!\!\!\!\!\!\!\!\!\!\!\!\!\!\!  + \frac{1}{2} \left[ \frac{\partial^2 C(0,0)}{\partial x^2}  +\left(  \frac{\sigma}{2} \frac{k_0  l'^2}{E I_y} \right) \frac{\partial C(0,0)}{\partial y} \right]\, x\,  V_G^2 \nonumber \\
&  & \!\!\!\!\!\!\!\!\!\!\!\!\!\!\!\!\!\!\!\!\!\!\!\!\!\!\!\!\!  + \frac{1}{2} \left[ \frac{1}{2} \frac{\partial^3 C(0,0)}{\partial x^3}+ \left( \frac{\sigma}{2} \frac{k_0  l'^2}{E I_y} - \frac{3}{5}\frac{1}{l'}\right) \frac{\partial^2 C(0,0)}{\partial x \partial y}  \right]\, x^2\,  V_G^2 \nonumber \\
&  & \!\!\!\!\!\!\!\!\!\!\!\!\!\!\!\!\!\!\!\!\!\!\!\!\!\! + \frac{1}{2} \left[ \frac{1}{6} \frac{\partial^4 C(0,0)}{\partial x^4} + \left( \frac{\sigma}{4} \frac{k_0  l'^2}{E I_y} - \frac{3}{5}\frac{1}{l'}\right) \frac{\partial^3 C(0,0)}{\partial x^2 \partial y} \right. \nonumber \\
& & \left.  \!\!\!\!\!\!\!\!\!  -\left( \frac{\sigma}{2} \frac{k_0  l'^2}{E I_y} \frac{3}{5}\frac{1}{l'} \right)  \frac{\partial^2 C(0,0)}{\partial y^2} \right]\, x^3\,  V_G^2 \label{capa}.
\end{eqnarray}
\begin{figure}[t!]
\includegraphics[height=6.5 cm]{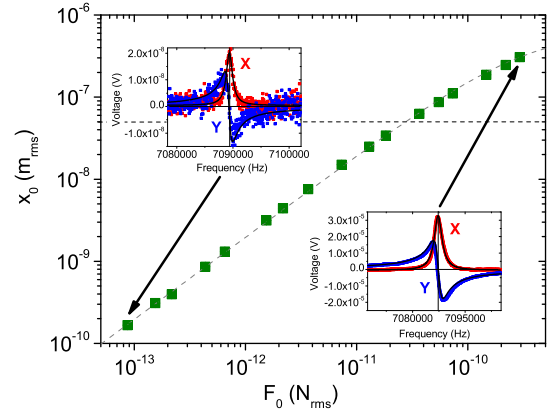}
\caption{\label{XvsF} (Color online) 
Main graph: NEMS response to a magnetomotive drive (field of 840$~$mT, no gate voltage). The dashed line is explained in the text. Note the wide dynamic range available, with the dashed horizontal representing the upper $x$ limit on the parametric amplification experiment. Error bars about $\pm 5~$\%, size of symbols. Insets: Actual resonance lines obtained at the two extremities of the graph (lines are Lorentzian fits).  }
\end{figure}
From the above equation, we immediately realize that the applied voltage $V_G$ allows to:
\begin{itemize}
\item drive the NEMS, with the first term,
\item tune the resonance frequency, with the second term,
\item and adjust the Duffing-like nonlinearity, with the last terms.
\end{itemize} 
These NEMS capacitive frequency tuning and nonlinearity tuning have been experimentally demonstrated for the first time in Ref. \cite{tuningRoukes}.
In the present work we use the resonance frequency tuning to realize the parametric drive \cite{rugarPRL,torsionalparpia}, with a proper choice of gate voltage modulation $V_G(t)$ \cite{qfs}.\\
The spatial variation of the capacitance $C(x,y)$ occurs on a typical lengthscale of the order of the electrode's gap $g$. Therefore, the terms in Eq. (\ref{capa}) between parenthesis should be compared to this parameter. We obtain $ \frac{\sigma}{2} \frac{k_0  l'^2}{E I_y}g \approx 0.01$ and $\frac{3}{5}\frac{1}{l'} g \approx 0.02$, and conclude that the corresponding factors in Eq. (\ref{capa}) can be safely neglected: only the $x$ variation of the NEMS capacitance $C$ has to be considered.

In practice, the clock and the two generators in Fig. \ref{setup} ($V_G$ and $V_I$) are implemented by a Tektronix AFG3252 dual channel arbitrary waveform generator.
The detection is realized with a Stanford SR844 RF lock-in amplifier, giving access to both the in-phase $X$ and out-of-phase components $Y$ of a harmonic motion $\dot{x}=x_0 \, d [ \cos (\omega t + \varphi) ]/dt$ (homodyne detection). The magnetic field is obtained with a small superconducting coil and a Kepco 10$~$A DC current source.

In Fig. \ref{XvsF} we present the characteristic displacement amplitude $x_0$ (at resonance) versus magnetomotive force $F_0$, obtained in a field of 840$~$mT (no voltage on the gate). The setup has been carefully calibrated to produce this curve \cite{qfs}, while preserving a high impedance environment for the NEMS.
To measure the resonance, the frequency $\omega$ of the excitation $F=F_0 \cos (\omega t)$ is swept around the mode frequency $\omega_0$ of the device. The smallest resonance line together with the biggest are shown in the insets, Fig. \ref{XvsF}. 

\begin{figure}[t!]
\includegraphics[height=6.5 cm]{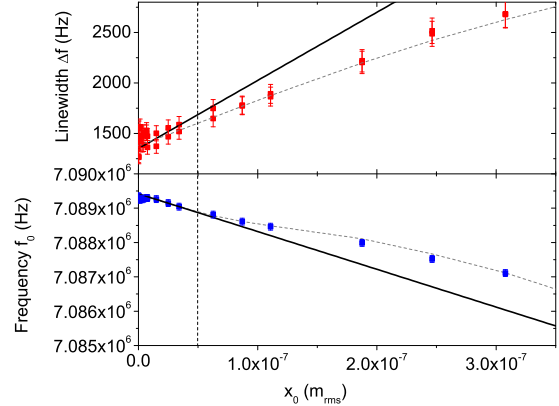}
\caption{\label{anel} (Color online) 
Top graph: Broadening of the line for large deflections $x_0$. Error bars about $\pm 5~$\%. Bottom graph: Frequency shift occuring in the same conditions. Error bars about $\pm 100~$Hz, about the size of symbols. On both graphs the full lines represent the linear-in-$x$ anelastic effect (top graph, slope $K=+5.\,10^6~$m$^{-1}$, and bottom graph slope $K'/2=-1.5\,10^3~$m$^{-1}$). The dashed curves fitting the data at large deflections arise from a thermal model taking also into account the Joule heating of the structure. The vertical dashed line is the upper $x$ limit on the parametric amplification experiment. }
\end{figure}

The full lines are Lorentzian fits demonstrating the exceptional linearity of the device: the intrinsic Duffing-like nonlinearity is extremely small. In fact, a cantilever is inherently less nonlinear than a doubly clamped beam which elongates under flexure \cite{book3,qfs}. \\
For small displacements, Fig. \ref{XvsF} enables to characterize the sample yielding the mode spring constant $k_0=2.5~$N/m, and thus the mode mass $m_0=1.2\,10^{-15}~$kg. The quality factor $Q$ is about $5.\,10^3$.
However, a nonlinear damping is clearly visible for large deflections (slight curvature at the top end of the $x_0$ versus $F_0$ curve). In Fig. \ref{anel} we present both the corresponding frequency shift and line broadening experienced by the resonance under large distortions.
For not too large excitation amplitudes (currents in the metallic layer), the shift and broadening are solely due to the {\it anelasticity} of the soft aluminum coating, an effect which is present in both NEMS and MEMS covered with soft metals \cite{coatings}. 
The experimental evidence is that both the metal Young modulus $E$ (thus the spring constant $k_{foot}$, and in turn the frequency $\omega_0$) and the damping coefficient $\Lambda$ (thus the linewidth $\Delta \omega =2 \Lambda / m_0 $) depend almost linearly on the amplitude of the strain in the layer, leading in turn to a linear dependence on $x_0$ (full lines in Fig. \ref{anel}).
For larger deflections, the currents needed to generate the magnetomotive force $F_0$ heat the structure (Joule effect in the 110$~\Omega$ layer; the mechanical heat dissipated per cycle remains negligible). This can be taken into account with a thermal model of the beams, leading to the dashed lines in Figs. \ref{anel} and \ref{XvsF}. 
In a separate publication we show that the thermal effects can be accurately described for temperatures ranging from 1.5$~$K to 30$~$K, with drive currents up to 50$~\mu$A \cite{JAPtobe}. 
 
For the present experiment, we keep the displacements $x$ below 50$~$nm. The heating of the structure is always negligible (even with the parametric drive on), and the dynamics equation describing the motion (with no gate voltage) is:
\begin{equation}
m_0 \ddot{x}+ 2 \Lambda(x) \dot{x} +2 k_{foot}(x) x = F(t). \label{eq}
\end{equation}
At first order, we will consider that $k_{foot}(x)$ and $\Lambda(x)$ are linear functions of the amplitude $x_0$ of the motion \cite{JLTP_VIW, coatings}:
\begin{eqnarray}
2 k_{foot}(x_0) & = & k_{0} \left( 1+ K' x_0 \right) \label{kx}  , \label{ko} \\
\Lambda(x_0) & = & \Lambda_0 \left( 1+ K x_0 \right) \label{lx} . \label{lambda}
\end{eqnarray}
For magnetic fields $B$ smaller than 1$~$T and voltages $V_G$ on the gate smaller than 8$~$V, no anomalies on the dynamic properties of the NEMS could be detected.

\begin{figure}[t!]
\includegraphics[height=6 cm]{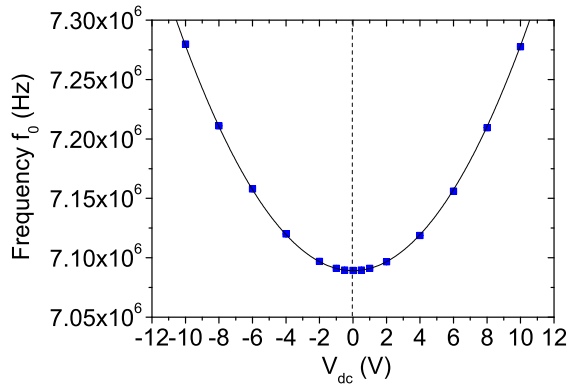}
\caption{\label{tune} (Color online) 
Demonstration of the frequency tuning ability of the device, with a DC voltage applied on the gate. The displacement $x$ has been kept small enough to be in the linear regime. We obtain $-\partial^2 C / \partial x^2 \approx 0.0027~$F/m$^2$.}
\end{figure}

In Fig. \ref{tune} we present the frequency tuning of our device. A DC voltage $V_G$ is applied on the gate electrode, leading to the expected quadratic dependence (black line).
In these measurements, we keep the displacement $x$ smaller than 7$~$nm, in order to ensure that the nonlinear terms in Eq. (\ref{capa}) remain negligible.
 From the fit, we obtain the spring constant tuning factor $-\partial^2 C / \partial x^2 \approx 0.0027~$F/m$^2$ \cite{qfs}.
The tunability is of the order of 200$~$kHz per 10$~$V and positive, which is in agreement with the literature
for out-of-plane motion \cite{tuningRoukes}. Note that this value is about a factor of 4 larger than the one reported in the preceding reference, since our electrode gap is 4 times smaller (100$~$nm). \\
With a nonzero voltage bias and larger deflections $x$, the device becomes nonlinear \cite{qfs}. A careful study of the dynamics enables to extract all the relevant terms appearing in Eq. (\ref{capa}) \cite{JAPtobe}.

\section{THEORETICAL MODELING}
\label{theory}

We consider a {\it single degree of freedom} mechanical structure that is driven into harmonic motion close to one of its resonance modes. 
For our purpose, it shall be a cantilevered beam in its out-of-plane flexure. 
Geometrical nonlinearities have been discussed in great details in Ref. \cite{nonlinPRB}. Materials nonlinearities (anelasticity) in aluminum coated devices have been studied in Ref. \cite{coatings}. From these works, we write down the most generic dynamics equation describing our devices, {\it without} any voltage on the gate electrode:
\begin{eqnarray}
  m_0 \left(1+ m_1 \, x + m_2 \, x^2 \right) \ddot{x}  &+ & \nonumber \\
  \left( \Gamma_0+ \Gamma_1 \, x \right)\, \dot{x}^2   & +& \nonumber \\ 
 2 \Lambda(x_0)  \left( 1+ l_1 \, x + l_2 \, x^2\right)\, \dot{x} & +& \nonumber  \\ 
 2 k_{foot}(x_0) \left( 1+ k_1 \, x + k_2 \, x^2\right) \!\!\! &  x &  \nonumber \\ 
 & = & \!\!  F_0 \cos (\omega t +\phi).  \label{nonlin}
\end{eqnarray}
The above equation contains both inertia nonlinear terms ($m_1,m_2$) and a nonlinear restoring force (with $k_1,k_2$). 
For our oscillators, the anelasticity appears both in the spring constant $k_{foot}(x_0)$ and the damping term $\Lambda(x_0)$, with $x_0$ the amplitude of the motion at frequency $\omega$. The $l_1,l_2$ and $\Gamma_0,\Gamma_1$ factors are also a consequence of geometrical nonlinearities, but could of course incorporate peculiar damping terms in the most general formalism \cite{nonlinPRB}. Eq. (\ref{eq}) appears to be the limit of Eq. (\ref{nonlin}) when geometrical nonlinearities are negligible. For the sake of completeness, we keep {\it all the nonlinear terms} in the present modeling. The aim is to establish theoretical {\it analytic} solutions which are as generic as possible, and usable for a broad range of micro/nano mechanical devices. These solutions are then illustrated with our experiments in the two following sections, for the {\it linear and nonlinear} regimes respectively.

Take Eq. (\ref{capa}) with a gate modulation $V_G = V_{dc}+ V_0 \cos (\omega' t + \phi')$. 
Since the force involves $V_G^2$, one could in principle chose $\omega'= \omega$ for the degenerate parametric drive. However, the capacitive coupling will also impose a current $I_C = - V_0 (\omega' C_0) \sin (\omega' t + \phi')$ through the NEMS, which will add up to the standard drive current $I$. As a result, this choice will spoil the magnetomotive test force $F$ to be amplified. Similarly, with a piezoelectric drive, a gate voltage $V_G$ at frequency $\omega$ will necessarily induce a parasitic motion of the oscillator.\\
We thus have to chose $\omega'= 2 \omega$, obtaining:
\begin{displaymath}
\!\! V_G^2 =  V_{dc}^2+ \frac{V_0^2}{2}  + 2 V_{dc} V_0 \cos (2 \omega t + \phi') +  \frac{V_0^2}{2} \cos (4 \omega t + 2 \phi').
\end{displaymath}
The capacitive force that has to be added to Eq. (\ref{nonlin}) writes then:
\begin{eqnarray}
F_C^{ef\!f.} &= & F_{dc} + F_{ac} \left[ \cos (2 \omega t + \phi') +  \alpha \cos (4 \omega t + 2 \phi') \right] \nonumber \\
&\!\!\!\!\!\!\!\!\!\!\!\!\!\!\!\!\!\!\!\!\!\!\!\!\!\!\!\!\!\!\!\!\!\!\! - &\!\!\!\!\!\!\!\!\!\!\!\!\!\!\!\!\!\!\!\!  k_0' x - \Delta k_0 x \, \left[ \cos (2 \omega t + \phi') +  \alpha \cos (4 \omega t + 2 \phi') \right] \nonumber \\
&\!\!\!\!\!\!\!\!\!\!\!\!\!\!\!\!\!\!\!\!\!\!\!\!\!\!\!\!\!\!\!\!\!\!\! - &\!\!\!\!\!\!\!\!\!\!\!\!\!\!\!\!\!\!\!\!  k_1' x^2 - \Delta k_1 x^2 \, \left[ \cos (2 \omega t + \phi') +  \alpha \cos (4 \omega t + 2 \phi') \right] \nonumber \\
&\!\!\!\!\!\!\!\!\!\!\!\!\!\!\!\!\!\!\!\!\!\!\!\!\!\!\!\!\!\!\!\!\!\!\! - &\!\!\!\!\!\!\!\!\!\!\!\!\!\!\!\!\!\!\!\!  k_2' x^3 - \Delta k_2 x^3 \, \left[ \cos (2 \omega t + \phi') +  \alpha \cos (4 \omega t + 2 \phi') \right]. \label{pump}
\end{eqnarray}
We have defined, from Eq. (\ref{capa}):
\begin{eqnarray*}
F_{dc}      & = & +\frac{1}{2} \frac{\partial C(0,0)}{\partial x}  \left( V_{dc}^2+ \frac{V_0^2}{2} \right) , \\
F_{ac}      & = & +\frac{1}{2} \frac{\partial C(0,0)}{\partial x}  \left( 2 V_{dc} V_0  \right) , \\
k_0'        & = & -\frac{1}{2} \frac{\partial^2 C(0,0)}{\partial x^2} \left( V_{dc}^2+ \frac{V_0^2}{2} \right) , \\
\Delta k_0  & = & -\frac{1}{2} \frac{\partial^2 C(0,0)}{\partial x^2} \left( 2 V_{dc} V_0  \right), \\
k_1'        & = & -\frac{1}{4} \frac{\partial^3 C(0,0)}{\partial x^3} \left( V_{dc}^2+ \frac{V_0^2}{2} \right) , \\
\Delta k_1  & = & -\frac{1}{4} \frac{\partial^3 C(0,0)}{\partial x^3} \left( 2 V_{dc} V_0  \right), \\
k_2'        & = & -\frac{1}{12} \frac{\partial^4 C(0,0)}{\partial x^4} \left( V_{dc}^2+ \frac{V_0^2}{2} \right) , \\
\Delta k_2  & = & -\frac{1}{12} \frac{\partial^4 C(0,0)}{\partial x^4} \left( 2 V_{dc} V_0  \right).
\end{eqnarray*}
We introduced $\alpha= V_0/(4 V_{dc})$ which quantifies the "contamination" of the parametric pumping by $4\omega$ harmonics. In our experiment, $\alpha < 1$.

Combining Eq. (\ref{nonlin}) and Eq. (\ref{pump}) brings finally:
\begin{eqnarray}
  m_0 \left(1+ m_1 \, x + m_2 \, x^2 \right) \ddot{x}  +  \left( \Gamma_0+ \Gamma_1 \, x \right)\, \dot{x}^2  & &\nonumber \\
 + 2 \Lambda(x_0)  \left( 1+ l_1 \, x + l_2 \, x^2\right)\, \dot{x} & & \nonumber  \\ 
 + \left( 2 k_{foot}(x_0)+k_0' \right) \left( 1+ \tilde{k}_1 \, x + \tilde{k}_2 \, x^2\right) \!x &   &  \nonumber \\ 
 + \Delta k_0 x \, \left[ \cos (2 \omega t + \phi') +  \alpha \cos (4 \omega t + 2 \phi') \right] & &  \nonumber \\ 
 + \Delta k_1 x^2 \, \left[ \cos (2 \omega t + \phi') +  \alpha \cos (4 \omega t + 2 \phi') \right] & &  \nonumber \\
 + \Delta k_2 x^3 \, \left[ \cos (2 \omega t + \phi') +  \alpha \cos (4 \omega t + 2 \phi') \right] & &  \nonumber \\
  =    F_{dc} + F_0 \cos (\omega t +\phi) &  & \nonumber \\ 
 + F_{ac} \left[ \cos (2 \omega t + \phi') +  \alpha \cos (4 \omega t + 2 \phi') \right] .& &   \label{equation}
\end{eqnarray}
The only non-explicit time dependence is in $x(t)$. We used the notations $\tilde{k}_1=(2 k_{foot} \, k_1+k_1')/(2 k_{foot} + k_0')$ and $\tilde{k}_2=(2 k_{foot} \,k_2+k_2')/(2 k_{foot} + k_0')$. 
We introduce both the phase of the excitation force $\phi$ and of the pump $\phi'$, with respect to the motion $x(t)$ taken as our phase reference.
$\Delta k_0$ is the usual (linear) parametric modulation term.

As is clearly seen on Eq. (\ref{equation}), capacitive parametric pumping generates unavoidably a very rich nonlinear dynamics equation: the pumping {\it itself} is nonlinear (with $\Delta k_1, \Delta k_2$, $F_{ac}$, and $\alpha$). $F_{dc}$ is a static deflecting force. Note nonetheless that the modeling can be adapted to other types of parametric excitations, by equating to zero the unnecessary parameters.

In order to resolve Eq. (\ref{equation}), we apply the method described in Ref. \cite{nonlinPRB}. The calculation is based on an idea first introduced by Landau \& Lifshitz \cite{landau}.
In order for the experiment to be meaningful, $F_{dc}$ (static deflection force), $F_{ac}$ (second harmonic excitation) and $\alpha$ (4th order contamination) have to remain small.  
Mathematically, we will consider that $F_{dc}$ and $F_{ac}$ are of the same order as $x_0^2$ terms, and we will reduce the calculation to the $\alpha=0$ case.

We take for the solution:
\begin{displaymath}
x(t) = \sum_{n=0}^{+\infty} a^c_{n}(\omega) \cos (n\,\omega t) + \sum_{n=1}^{+\infty} a^s_{n}(\omega) \sin (n\,\omega t)
\end{displaymath}
and seek only the static term $n=0$, plus the first harmonic $n=1$.
The calculation requires also $a^c_{2}, a^s_{2}$ to be retained. We have $x_0=\sqrt{(a^c_{1})^2+(a^s_{1})^2}$ by definition.
The obtained amplitudes write:
\begin{widetext}
\vspace*{-8 mm}
\begin{eqnarray*}
a^c_{0} & = & \frac{F_{dc}}{ 2 k_{foot} + k_0'} + \beta_0 \, x_0^2 , \\
 a^c_{1}(\omega) &\!\!\!\!\! = &\!\!\!\!\! \frac{F_{0}}{m_0} \left[ \frac{(\omega_r^2-\omega^2) \cos(\phi) + \left(\Delta \omega \, \omega \right) \sin(\phi) - \Delta_{+} \cos(\phi-\phi') - \Delta_{-} \cos(\phi+\phi') + \delta_{+} \sin(\phi-\phi') - \delta_{-} \sin(\phi+\phi') }{\left(\omega_r^2-\omega^2 \right)^2+ \left(\Delta \omega \, \omega \right)^2- \left[ \Delta_{+}^2+\Delta_{-}^2+\delta_{+}^2+\delta_{-}^2 + 2 \left( \Delta_{+}\Delta_{-} - \delta_{+}\delta_{-} \right) \cos(2 \phi')  + 2 \left( \Delta_{+}\delta_{-} + \delta_{+}\Delta_{-} \right) \sin(2 \phi') \right]} \right], \\
a^s_{1}(\omega) &\!\!\!\!\! = &\!\!\!\!\! \frac{F_{0}}{m_0} \left[ \frac{\left(\Delta \omega \, \omega \right) \cos(\phi) + (\omega^2-\omega_r^2) \sin(\phi) - \Delta_{+} \sin(\phi-\phi') - \Delta_{-} \sin(\phi+\phi') - \delta_{+} \cos(\phi-\phi') + \delta_{-} \cos(\phi+\phi') }{\left(\omega_r^2-\omega^2 \right)^2+ \left(\Delta \omega \, \omega \right)^2- \left[ \Delta_{+}^2+\Delta_{-}^2+\delta_{+}^2+\delta_{-}^2 + 2 \left( \Delta_{+}\Delta_{-} - \delta_{+}\delta_{-} \right) \cos(2 \phi')  + 2 \left( \Delta_{+}\delta_{-} + \delta_{+}\Delta_{-} \right) \sin(2 \phi') \right]} \right] .
\end{eqnarray*}
\end{widetext}
If the phase reference is turned to match $\phi$, the pumping terms write $\cos(2\phi \pm \phi'),\sin(2\phi \pm \phi')$ while the $\sin(\phi)$ disappears and the $\cos(\phi)$ becomes 1. The cosine and sine appearing in the denominator change to $\cos(2 \phi'- 4 \phi), \sin(2 \phi'- 4 \phi)$ respectively.
In these expressions, the resonance position $\omega_r$, the linewidth $\Delta \omega$ and the pumping terms $\Delta_{+},\Delta_{-},\delta_{+},\delta_{-}$ {\it are themselves functions} of the amplitudes $ a^c_{1}, a^s_{1}$:
\begin{eqnarray*}
\omega_r^2 & = & \omega_0^2+ 2 \omega_0 \, \delta \omega + 2 \beta_1 \omega_0 \, x_0^2 + 2 \omega_0 \beta_1'\, a^c_{1}  a^s_{1} , \,\,\, i.e. \\
\omega_r & \approx & \omega_0+ \delta \omega + \beta_1 \, x_0^2 + \beta_1'\, a^c_{1}  a^s_{1},\\
\Delta \omega & = & \Delta \omega_0 + \delta (\Delta \omega) + \beta_2 \, x_0^2 + \beta_2'\, a^c_{1}  a^s_{1} ,\\
\Delta_{+} & =& \Delta_{+}^0 + \Gamma_{+} \, x_0^2, \\
\Delta_{-} & =& \Gamma_{-} \, x_0^2, \\
\delta_{+} & =& \delta_{+}^0 + \gamma_{+} \, x_0^2, \\
\delta_{-} & =& \gamma_{-} \, x_0^2, 
\end{eqnarray*}
with the usual definitions $\omega_0=\sqrt{(2 k_{foot}+k_0')/m_0}$, and $\Delta \omega_0 =2 \Lambda/m_0$ (expressed in Rad/s). 
The above is valid for {\it any } $Q=\omega_0/\Delta \omega$.

The nonlinear parameters are given in the Appendix \ref{appendix} as a function of the terms appearing in Eq. (\ref{equation}). 
Note that $k_{foot}$ and $\Lambda$ shall not depend on the (vanishingly small, tens of pm in our experiments) static deflection induced by $F_{dc}$ and the nonlinear coefficient $\beta_0$.
Note also that mathematically the spurious forces $F_{dc}, F_{ac}$ combined to the nonlinear first order terms $m_1,\tilde{k}_1$, and to the pumping coefficients $\Delta k_0,\Delta k_1$ induce a resonance shift $\delta \omega$. 
The peculiar pumping term $\delta_{+}^0$ appears from a similar combination. 
The static force $F_{dc}$ with $l_1$ induce an additional damping term $\delta (\Delta \omega)$ as well. In practice, these terms appear to be  completely negligible for our devices (see Appendix \ref{appendix}). \\
In the high $Q$ limit, the writing of the nonlinear parameters simplify considerably (see Appendix \ref{appendix}), and in particular $\delta{+},\delta_{-} \approx 0$.
When all the pumping terms are taken to be zero in Eq. (\ref{equation}), the above expressions reduce to the result given in Ref. \cite{nonlinPRB}. The {\it standard} parametric amplification theory \cite{book2,book3} considers all nonlinear terms to be zero, with only the pumping term $\Delta k_0 \neq 0$. In this particular case $a^c_{0}=0$, the terms $\delta \omega, \delta (\Delta \omega)=0$, the nonlinear coefficients $\beta_1,  \beta_1', \beta_2,  \beta_2'=0$ and the only pumping factor that remains is $\Delta_{+}^0=\Delta k_0/(2 m_0)$, as it should.
\begin{figure}[t!]
\includegraphics[height=5 cm]{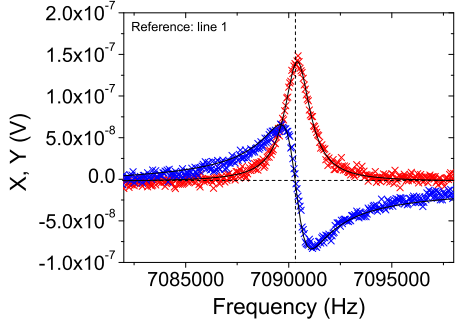}
\caption{\label{ref} (Color online) 
Reference line for the phase dependence of the gain (parametric pumping $h=0$). The field was 840$~$mT and the current $110~$nA, thus the force $660~$fN. The maximal displacement is $1.4~$nm (rms values). The black lines are Lorentzian fits, giving access to the resonance frequency $7.090~$MHz and linewidth $1490~$Hz.}
\end{figure}

In the experimental following sections, we will be concerned with  $a^c_{1}, a^s_{1}$ in the simple case $\phi=0$, and $\phi'=\pm \pi/2$.
The expressions reduce then to:
\begin{eqnarray}
\!\!\!\!\!\!\!\!\!\!\!\!\!\! a^c_{1}(\omega) & = & \frac{F_{0}}{m_0} \, \times \nonumber \\ 
 & &\!\!\!\!\!\!\!\!\!\!\!\!\!\!\!\!\!\!\!\!\!\!\!\!\!\!\! \left[ \frac{(\omega_r^2-\omega^2) -(\pm 1) \left( \delta_{+} + \delta_{-} \right) }{\left(\omega_r^2-\omega^2 \right)^2+ \left(\Delta \omega \, \omega \right)^2- \left( \Delta_{+}-\Delta_{-} \right)^2 - \left( \delta_{+}+\delta_{-} \right)^2} \right], \label{Xpump} \\
\!\!\!\!\!\!\!\!\!\!\!\!\!\! a^s_{1}(\omega) & = & \frac{F_{0}}{m_0}  \, \times \nonumber \\
 & &\!\!\!\!\!\!\!\!\!\!\!\!\!\!\!\!\!\!\!\!\!\!\!\!\!\!\! \left[ \frac{\left(\Delta \omega \, \omega \right)  +(\pm 1) \left( \Delta_{+}  - \Delta_{-} \right)  }{\left(\omega_r^2-\omega^2 \right)^2+ \left(\Delta \omega \, \omega \right)^2 - \left( \Delta_{+}-\Delta_{-} \right)^2- \left( \delta_{+}+\delta_{-} \right)^2 } \right] \label{Ypump} .
\end{eqnarray}
The sign $\pm$ corresponds to the sign of $\phi'=\pm \pi/2$: it is $+$ for amplification and $-$ for squeezing. These expressions are valid {\it even in} the nonlinear case, for {\it any} $Q$.

We will use a generalized {\it nonlinear} definition of the pumping factor $h$:
\begin{eqnarray}
h(x_0) & = & (\pm 1) \left( \frac{\Delta_{+} -\Delta_{-} }{\Delta \omega \, \omega} \right). \label{hfact}
\end{eqnarray}
This number is $\left| h \right| < 1$ in the parametric amplification regime under study, positive for amplifying and negative for squeezing.

In the next section, we will also be interested in the $\phi,\phi'$ angular dependence at resonance $\omega=\omega_r$ of the amplitude $x_0=\sqrt{(a^c_{1})^2+(a^s_{1})^2}$, in the limit of small displacements. We obtain:
\begin{eqnarray}
 x_0(\omega_r) & = & \frac{F_{0}}{m_0} \, \times \nonumber \\ 
& &\!\!\!\!\!\!\!\!\!\!\!\!\!\! \left[ \left( \Delta \omega \, \omega_r - \Delta_{+} \right)^2 \sin^2 \left(\frac{\pi}{4} +  \frac{2\phi -\phi'}{2}\right) \right. \nonumber \\
& &\!\!\!\!\!\!\!\!\!\!\!\!\!\! \left. + \left( \Delta \omega \, \omega_r + \Delta_{+} \right)^2 \cos^2 \left(\frac{\pi}{4} +  \frac{2\phi -\phi'}{2}\right) \right. \nonumber \\
& &\!\!\!\!\!\!\!\!\!\!\!\!\!\!\!\!\!\!\!\!\!\!\!\!\!\!\!\!\!\!\!\!\!\!\!\!\!\! \left.  + \delta_{+}^2 - 4\delta_{+} \, \Delta \omega \, \omega_r \cos \left(\frac{\pi}{4} +  \frac{2\phi -\phi'}{2}\right) \sin \left(\frac{\pi}{4} +  \frac{2\phi -\phi'}{2}\right)\right]^{1/2} \nonumber \\
& & / \left[ \left(\Delta \omega \, \omega_r \right)^2 -\left( \Delta_{+}^2 + \delta_{+}^2 \right) \right] \label{Phase},
\end{eqnarray}
valid in the linear regime ($\Gamma_{\pm}$, $\gamma_{\pm}$ and all $\beta$ terms taken to zero), for {\it any} $Q$.

Equations Eqs. (\ref{Xpump} - \ref{Phase}) will be used and commented in Sections \ref{linear},\ref{nonlinear}; practical limits on parametric amplification
 will be discussed and illustrated in Section \ref{pratic} on their basis.

\section{LINEAR PARAMETRIC AMPLIFICATION}
\label{linear}

\begin{figure}[t!]
\includegraphics[height=6 cm]{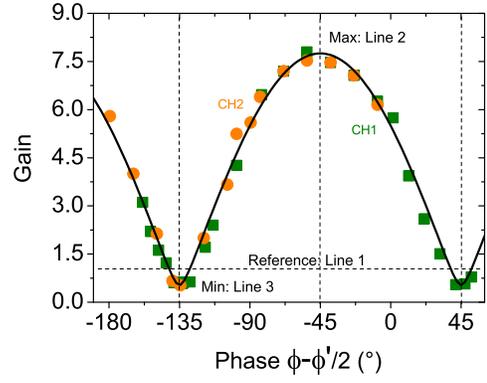}
\caption{\label{angle} (Color online) 
Phase dependence of the parametric gain with respect to $\phi-\phi'/2$, obtained from the height of the resonance peak at $\omega=\omega_r$. The reference line is shown in Fig. \ref{ref}. The lines obtained at the maximum and minimum are shown in Figs. \ref{max} and \ref{min} respectively. The dark (green, CH1) squares are obtained when tuning the phase of the force $\phi$. The light (orange, CH2) dots are recalculated when adjusting the parametric modulation angle $\phi'$. The black line is the fit explained in the text, with a pumping factor of $\left|h\right|=0.875$.}
\end{figure}

\begin{figure}[t!]
\includegraphics[height=5 cm]{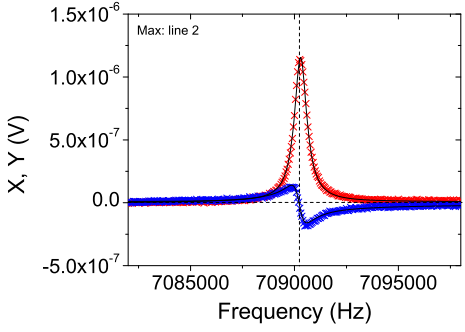}
\caption{\label{max} (Color online) 
Same settings as Fig. \ref{ref} (reference line) with the pump on ($\left|h\right|=0.875$). The phase is set to be on the maximum of the gain amplification, Fig. \ref{angle}. Black lines are Lorentzian fits explained in the text, demonstrating the assymetry between the two quadratures. The fit linewidth $\Delta f_{pump}$ is of $725~$Hz, for a gain of $8$.}
\end{figure}

\begin{figure}[t!]
\includegraphics[height=5 cm]{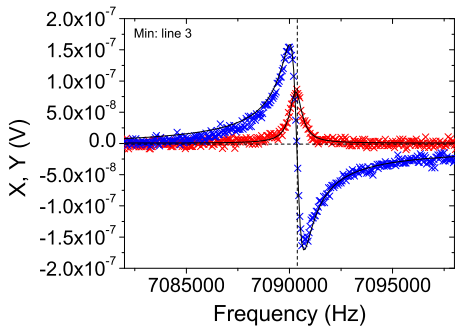}
\caption{\label{min} (Color online) 
Same settings as Fig. \ref{ref} (reference line) with the pump on ($\left|h\right|=0.875$). The phase is set to be on the minimum of the gain amplification, Fig. \ref{angle}. Black lines are Lorentzian fits explained in the text, demonstrating the assymetry between the two quadratures. The fit linewidth $\Delta f_{pump}$ is of $710~$Hz, for a squeezing of $0.53$.}
\end{figure}

In Fig. \ref{angle} we present the phase dependence of the parametric gain measured for the device shown in Fig. \ref{setup}. 
The reference line, obtained when the pumping is turned off, is shown in Fig. \ref{ref}. 
Two particular positions are marked: the gain maximum (denoted line 2, shown in Fig. \ref{max}), and the gain minimum (squeezing, line 3 presented in Fig. \ref{min}). What is drawn is the amplitude of the detected voltage, proportional to the velocity $\dot{x}(t)$.

In our devices, $Q>\!>1$ and the $\delta_{+},\delta_{-}$ terms introduced in Section \ref{theory} can be safely neglected. Taking Eqs. (\ref{Xpump}) and (\ref{Ypump}), and injecting the definition of $h$, Eq. (\ref{hfact}), we obtain:
\begin{eqnarray}
a^c_{1}(\omega) & = & \!\! \frac{F_{0}}{m_0} \, \left[ \frac{(\omega_r^2-\omega^2)}{\left(\omega_r^2-\omega^2 \right)^2+ \left(\Delta \omega \, \omega \right)^2 \left( 1 -  h^2 \right) } \right] \!, \label{Xpumph} \\
 a^s_{1}(\omega) & = & \frac{F_{0}}{m_0} \, \sqrt{\frac{1+h}{1-h}} \, \times \nonumber \\
 & &  \left[ \frac{\left(\Delta \omega \, \omega \right) \sqrt{ 1 - h^2 } }{\left(\omega_r^2-\omega^2 \right)^2+ \left(\Delta \omega \, \omega \right)^2 \left( 1 -  h^2\right) } \right] \label{Ypumph} ,
\end{eqnarray}
written here for the linear regime (nonlinear terms of Section \ref{theory} taken to zero).
The functions in brackets {\it are still} Lorentzian expressions, with a linewidth decreased to $\Delta \omega  \sqrt{ 1 - h^2 } $ for {\it both} amplification and squeezing. Also, the parametric pumping has introduced an assymetry between the heights of the $X$ and $Y$ components with the factor $\sqrt{(1+h)/(1-h)}$: for amplification, one of the quadratures becomes large as $h \rightarrow 1$ while for squeezing, when $h \rightarrow -1$ it tends to $1/2$ of its nominal value \cite{rugarPRL,book2,book3}. These fits are illustrated in Fig. \ref{max} (amplification) and Fig. \ref{min} (squeezing) on two measured resonance lines.

The phase dependence Eq. (\ref{Phase}) can be recast:
\begin{eqnarray}
 x_0(\omega_r) & = & \frac{F_{0}}{m_0 \, \Delta \omega \, \omega_r} \, \times \nonumber \\
 & &\!\!\!\!\!\!\!\!\!\!\!\!\!\!\! \sqrt{ \frac{\sin^2 \left(\frac{\pi}{4} +  \frac{2\phi -\phi'}{2}\right)}{ \left( 1+ \left| h \right| \right)^2 } + \frac{\cos^2 \left(\frac{\pi}{4} +  \frac{2\phi -\phi'}{2}\right)}{ \left( 1- \left| h \right| \right)^2 } } , \label{Phaseh}
\end{eqnarray}
which contains the well-known gain function drawn on the data, Fig. \ref{angle} (i.e. the part under the square root) \cite{rugarPRL,book2,book3}.

\begin{figure}[t!]
\includegraphics[height=6.5 cm]{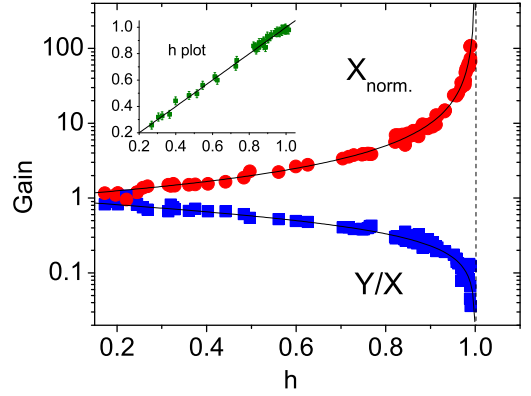}
\caption{\label{gainmax} (Color online) 
Parametric gain in amplification conditions as a function of the pumping $h$. The graph has been obtained in various magnetic fields, current drives and voltage biases conditions. All data taken with displacements kept below $15~$nm (rms). Note the log. scale. We plot both the height of the $X$ component normalized to the unpumped result, and the ratio between the height of the $Y$ to the $X$ component. Lines are fits explained in the text. Inset: comparison between the $h$ factor obtained from the fit linewidth, and the calculated one from the biases $V_{dc},V_0$. The maximal gain attained is $105$.}
\end{figure}

\begin{figure}[t!]
\includegraphics[height=6.5 cm]{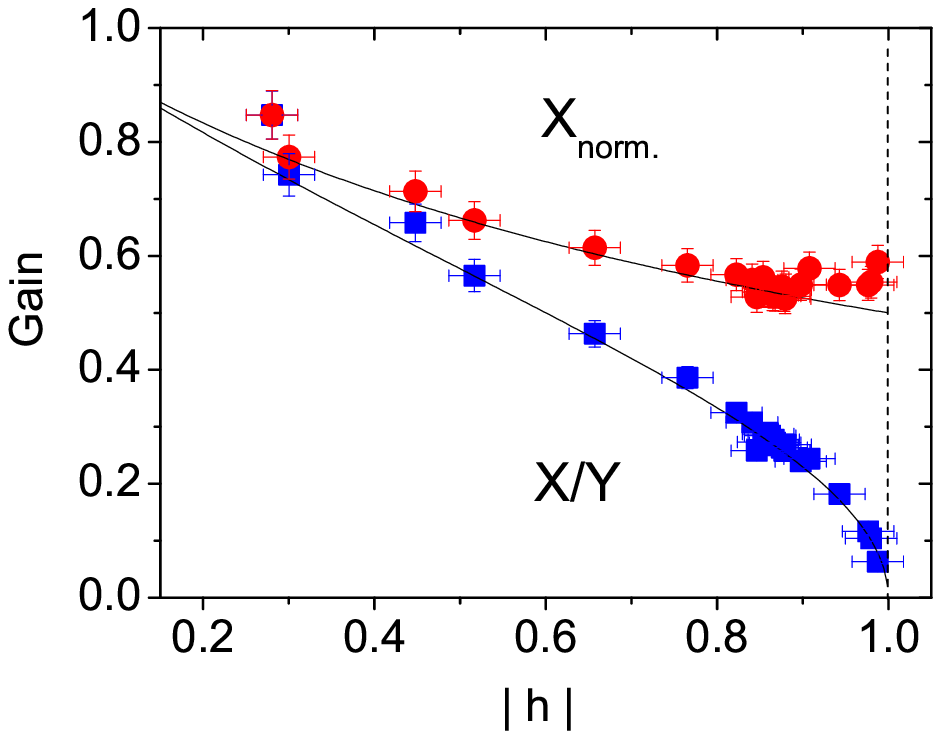}
\caption{\label{gainmin} (Color online) 
Parametric gain in squeezing conditions as a function of the pumping $\left|h\right|$. The graph has been obtained in various magnetic fields, current drives and voltage biases conditions. All data taken with displacements kept below $15~$nm (rms). We plot both the height of the $X$ component normalized to the unpumped result, and the ratio between the height of the $X$ to the $Y$ component. Lines are fits explained in the text. The strongest squeezing is $0.53$.}
\end{figure}

In Fig. \ref{gainmax} we show the parametric gain as a function of the pumping factor $h$ in amplifying conditions. In Fig. \ref{gainmin} we present the same data obtained in squeezing conditions. For both graphs, the gain is obtained from the height of the $X$ component at resonance $\omega=\omega_r$. From Eq. (\ref{Ypumph}) we immediately realize that it writes $1/(1-h)$ \cite{rugarPRL,book2,book3}; this simple fit is shown on both Figs. \ref{gainmax} and \ref{gainmin} (on $X_{norm.}$ data). Moreover, the ratio between the heights of the two quadratures (maximum at resonance for $X$, and twice the half-height at $\omega_r \pm \Delta\omega/2$ for $Y$) is $\sqrt{(1-\left|h\right|)/(1+\left|h\right|)}$ when taking for the numerator the largest of the two depending on amplifying/squeezing (see Eqs. (\ref{Xpumph}) and (\ref{Ypumph}) above). This fit is shown in Figs.  \ref{gainmax} and  \ref{gainmin} as well (on $Y/X$ and $X/Y$ data, respectively).

In all experiments, the pumping $h$ is realized by tuning the DC voltage $V_{dc}$ and the AC amplitude $V_0$ on the gate electrode. However, it can be measured {\it in situ} from the fit $\Delta \omega_{pump}$ of the resonance linewidth, knowing the "unpumped" value $\Delta \omega$: $\left| h \right|=\sqrt{1 - \left( \Delta \omega_{pump}/\Delta \omega \right)^2}$. This is precisely how the graphs presented in this section have been realized. We calculate the {\it imposed pumping} $h$ from Section \ref{theory}, keeping only the lowest order terms (see Appendix \ref{appendix}): $h=\frac{1}{2} Q (-\partial^2 C/\partial x^2)(V_{dc} V_0)/k_0$, with $2 k_{foot}+k_0' \approx k_0$. The coupling term $-\partial^2 C/\partial x^2$ is obtained from the frequency tuning graph, Fig. \ref{tune}. The resulting comparison between imposed pumping and measured pumping is given in the inset of Fig. \ref{gainmax}, proving excellent agreement. The voltages never exceeded $\left|V_{dc}\right|< 0.6~$V and $V_{0}< 0.7~$V$_{rms}$.

In these experiments, we demonstrate a very good agreement with theory. Forces down to about $6.5~$fN$_{rms}$ have been successfully amplified and detected with the present setup sensitivity: displacements down to about 0.15$~$nm$_{rms}$ in a bandwidth of 0.8$~$Hz around 7$~$MHz, with a (room temperature) amplifier noise floor of about 4$~$nV$_{rms}$/$\sqrt{\mathrm{Hz}}$. We show both squeezing and amplification. We report on a maximal gain of $105$, for a corresponding displacement of 1.7$~$nm$_{rms}$, which is about $2.6~$\% of the thickness of the moving beam, peak-to-peak. This linear gain is higher than in most experiments \cite{rugarPRL,pumphighQ,gaasstress,torsionalparpia,parpiadisk,roukeshighfreq,schwabqubit}, and is definitively the highest reported for a beam moving by such a substantial fraction \cite{highgain}. However, for too large deflections, the gain falls down and eventually the device jumps to parametric oscillations. This will be discussed in the two following sections.

\section{NONLINEAR PARAMETRIC AMPLIFICATION}
\label{nonlinear}

When the deflection $x_0$ becomes large, the device enters the nonlinear regime. In order for the Taylor expansion of Eq. (\ref{capa}) to remain meaningful, we keep the displacement $x_0 < 50~$nm. In this section, we consider that the working point of the device remains always {\it below} the threshold to parametric oscillation.
 Eqs. (\ref{Xpumph}) and (\ref{Ypumph}) are {\it still valid}, but with a nonlinear pumping factor $h$, Eq. (\ref{hfact}), and a nonlinear frequency $\omega_r$ and width $\Delta \omega$ (Section \ref{theory}).
Three sources of nonlinearities have to be taken into account: 
a constitutive geometrical nonlinearity giving rise to the $\beta$ terms,  
the nonlinear capacitive coupling used for the parametric pumping which generates the $\Gamma_{\pm}$, $\gamma_{\pm}$ and $\beta'$ terms, and finally a materials nonlinearity that we chose to describe experimentally with the two parameters $K$ and $K'$, Eqs. (\ref{ko}) and (\ref{lambda}). 
These last equations are not rigorously derived from a microscopic theory, but they reproduce what is seen experimentally in Section \ref{setups}, Fig. \ref{anel} \cite{coatings}.

\begin{figure}[t!]
\includegraphics[height=6.5 cm]{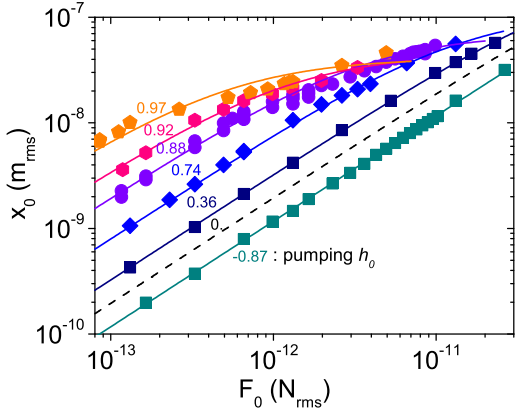}
\caption{\label{Fnonlin} (Color online) 
Displacement $x_0$ versus force $F_0$ curves, obtained for different parametric pumpings $h_0$ (see Fig.) in a field of 840$~$mT. The dashed line is the non-pumping $h_0=0$ reference, Fig. \ref{XvsF}, while the lines are fits explained in the text.}
\end{figure}

It is instructive to first look at the characteristic displacement $x_0$ versus force $F_0$ curve, for various pumping amplitudes, Fig. \ref{Fnonlin}.
From Eq. (\ref{Ypumph})  one easily obtains at first order and on resonance $\omega=\omega_r$:
\begin{eqnarray*}
\!\!\!\!\!\! x_0(\omega_r) & = & \frac{1}{(1-h_0)} \, \frac{F_{0}}{m_0 \, \Delta \omega_{00} \, \omega_{00}} \, \frac{1}{1+\frac{\left( K+\frac{K'}{2}\right)}{(1-h_0)^2} \frac{F_{0}}{m_0 \, \Delta \omega_{00} \, \omega_{00}}+...} 
\end{eqnarray*}
with $\Delta \omega_{00} = 2 \Lambda_0/m_0$, $\omega_{00}= \sqrt{k_0/m_0}$ and $h_0=\Delta_{+}^0/(\Delta \omega_{00} \, \omega_{00})$ the parameters obtained for $x_0 \rightarrow 0$, neglecting the small terms $\delta \omega$, $\delta (\Delta \omega)$ and $k_0'/k_0$ (see Appendix \ref{appendix}).  This expression is represented as lines on Fig. \ref{Fnonlin}, for the different $h_0$ pumpings shown.

\begin{figure}[t!]
\includegraphics[height=6.2 cm]{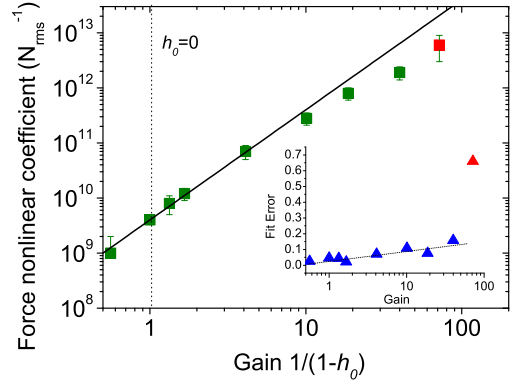}
\caption{\label{Amplinonlin} (Color online) 
Nonlinear parameter fit on Fig. \ref{Fnonlin} for different pumping amplitudes $h_0$. The line is a guide showing an $y=c\,x^2$ dependence. Inset: fit error for each curve adjusted on the data of Fig. \ref{Fnonlin}. The dashed line illustrates an increase tendency as one approaches $h_0 \rightarrow 1$; the last (masked) point is particularly erroneous, demonstrating the growing importance of higher-order nonlinear coefficients. }
\end{figure}

The key result is that the (purely material-dependent) nonlinear factor $\left( K+\frac{K'}{2}\right)$ affecting the top end of the curve in Fig. \ref{XvsF} is now {\it amplified} by a factor $1/(1-h_0)^2$, the gain squared. Similarly, the next order which includes geometrical and parametric nonlinearities is amplified by $1/(1-h_0)^3$. In practice, high-gain amplification is thus very sensitive to nonlinear parameters; on the other hand, squeezing {\it reduces} significantly the nonlinear behavior (Fig. \ref{Fnonlin}, \ref{Amplinonlin}), and can be proposed for this purpose.

Fitting the above expression and extracting the force nonlinear coefficient of the denominator, we obtain Fig. \ref{Amplinonlin}. The qualitative agreement is fairly good, with the proper power dependence. However, this modeling is too crude for a quantitative agreement: the original Taylor series involves $x$ and not $F$, and higher order terms in the force expansion (the $+...$ above) shall not be disregarded. 
In particular, as $h_0 \rightarrow 1$ (and the gain increases) the first-order expansion is clearly getting worse (with the amplification of all other nonlinear terms), and the last point fit in Fig. \ref{Amplinonlin} is of a particularly bad quality.

In order to obtain a quantitative understanding, we have to consider the corollary $h(x_0)$ function, Eq. (\ref{hfact}). From Section \ref{theory} we obtain, on resonance:
\begin{equation}
\!\!\!\!\!\!\!\!\! h(x_0)  =\frac{  h_0 \, \left[ 1 + \left( \frac{\Gamma_{+}-\Gamma_{-}}{\Delta_{+}^0}\right) x_0^2 \right]  }{ 1+ \left( K+\frac{K'}{2}\right) x_0 + \left( \frac{K\,K'}{2} + \frac{\beta_1}{\omega_{00}}+\frac{\beta_2}{\Delta \omega_{00}} \right) x_0^2  }. \label{hfit}
\end{equation}
The pumping is now {\it nonlinear}, i.e. it depends on the displacement $x_0$. Note that the $\beta'$ terms introduced in Section \ref{theory} disappear since on resonance the product $a^c_{1}  a^s_{1} $ is always zero.

In Fig. \ref{hnonlin} we show the measured pumping factor obtained from the fit of the linewidth, as a function of the measured displacement obtained from the peakheight. 
As can be seen, the major nonlinear effect is an almost linear reduction of the pumping strength with $x_0$. The lines in Fig. \ref{hnonlin} are fits based on Eq. (\ref{hfit}) were we impose $ K+\frac{K'}{2} \approx +5.\,10^6~$m$^{-1}$ (from Fig. \ref{anel}, Eqs. (\ref{ko}) and (\ref{lambda}), Section \ref{setups}). The only remaining nonlinear parameter (at second order) is a single quadratic term that we adjust simultaneously on all data. 

In the Appendix \ref{appendix}, we give all the nonlinear terms and discuss their relative importance. In practice, for our devices $\Delta_{+}^0 \approx \Delta k_0/(2 m_0)$ and $\Gamma_{+}-\Gamma_{-} \approx \Delta k_2/(4 m_0)$. $\beta_1/\omega_{00}$ and $\beta_2/\Delta \omega_{00}$ can be safely neglected, 
ensuring that the resonance lines remain Lorentzian-looking.
 The quadratic nonlinear term appearing in Fig. \ref{hnonlin} reduces then to $(\Gamma_{+}-\Gamma_{-})/\Delta_{+}^0 \approx (\partial^4 C/\partial x^4)/(12\, \partial^2 C/ \partial x^2)$. We obtain $-5.\,10^{13}~$$\pm 3.\,10^{13}~$m$^{-2}$ which is in good agreement with the estimated values for the capacitance coefficients introduced in Eq. (\ref{capa}) \cite{JAPtobe}.

In these experiments, we explored the nonlinear regime of parametric amplification. The data are fit with the nonlinear theory of Section \ref{theory} providing a good understanding of the phenomena. In particular, we show that the amplification increases the nonlinear behavior of the devices, and we propose the squeezing as a means to {\it reduce} nonlinear properties of parametrically pumped oscillators. 
In the amplification regime, we demonstrate the importance of the nonlinearity in the pumping scheme {\it itself}.

\section{PRACTICAL APPLICATIONS}
\label{pratic}

Parametric amplification is usually thought as a linear technique increasing the force sensitivity, or leading to a better frequency resolution through a narrower resonance line. It can be also thought as a technique used in the nonlinear regime of soft materials in order to probe {\it anelastic properties}. 

\begin{figure}[t!]
\includegraphics[height=6.5 cm]{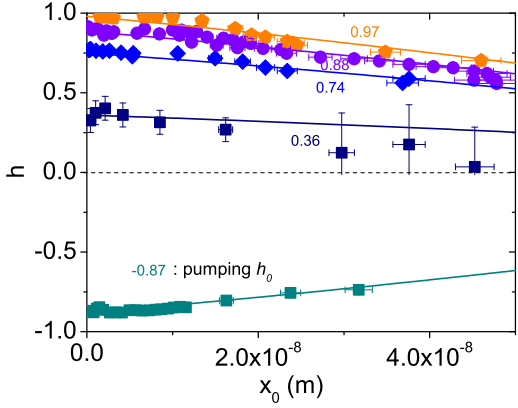}
\caption{\label{hnonlin} (Color online) 
Nonlinear pumping factor $h$ obtained from the fit linewidth $\Delta f_{pump}$ as a function of the displacement $x_0$. the $x_0 \rightarrow 0$ values $h_0$ are displayed on the graph. The lines are fits based on Eq. (\ref{hfit}) enabling the estimation of the pumping second order nonlinear term (see text).}
\end{figure}

\begin{figure}[t!]
\includegraphics[height=6.1 cm]{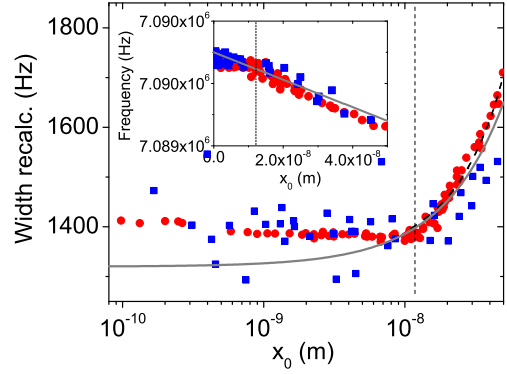}
\caption{\label{highRes} (Color online) 
Recalculated linewidth $\Delta f = 2 \pi \Delta \omega$ (red dots, various pumpings $h_0 \geq 0.75$) from Eq. (\ref{eqDWres}) compared to direct measurements (blue squares, no pumping). Note the log. scale. The original resolution was typically $\pm 5~$\% while the parametrically pumped result goes down to about $\pm 0.5~$\%. Inset: position of the resonance line. Dots (red) are the parametric pumping experiment (dispersion typ. $\pm 25~$Hz), while squares (blue) are direct measurements limited by the natural linewidth of the resonance (resolution of about 1/10th of $\Delta f$, $\pm 100~$Hz). The full lines are the empirical fits presented in Section \ref{setups}, Fig. \ref{anel}. The dashed line takes into account the nonlinear pumping coefficient $(\Gamma_{+}-\Gamma_{-})/\Delta_{+}^0$ (see text).}
\end{figure}

Neglecting the (second order) nonlinear $\beta$ and $\Gamma_{\pm}$ terms, and injecting Eq. (\ref{hfact}) in Eq. (\ref{Ypumph}) we obtain, for $\omega=\omega_r$:
\begin{equation}
\frac{F_0\,\omega_{00}}{x_0\, k_0}+\Delta \omega_{00} \, h_0 = \Delta \omega(x_0) , \label{eqDWres}
\end{equation}
which contains only materials nonlinearities. $\omega_{00}$, $\Delta \omega_{00}$ and $h_0$ are the zero-displacement values of the frequency, linewidth and pumping factor, respectively. The recalculated linewidth is shown in Fig. \ref{highRes} together with direct measurements. The plot has been obtained with various pumpings $h_0  \geq 0.75$. The inset shows the corresponding resonance position.

The increase on the linewidth resolution is impressive, reaching a value of about $\pm 0.5~$\%. On the other hand, the improvement of the resonance position precision is less marked. 
This result is perfectly normal: with i.e. a pumping factor of $h_0=0.98$, the gain which increases the recalculated linewidth resolution is about 50, while the line narrowing factor increasing the position precision is only of 5. Within the original error bars, position and linewidth seem to be almost linear with respect to $x_0$, Eqs. (\ref{ko}) and (\ref{lambda}) and full lines in Figs. \ref{highRes} and \ref{anel}. However, with our improved sensitivity we clearly distinguish a threshold effect: below about $12~$nm, the parameters seem to be constant, while only {\it above} they seem to increase linearly. We attribute this effect to the plastic limit of the aluminum coating of our device, which remains perfectly elastic below the limit and starts to show a strain-dependent behavior above. We estimate from Ref. \cite{coatings}, Eq. (27) that the plastic limit occurs at $\sigma_{Pl.} \approx 90~$ MPa, which is of the right order of magnitude.

For large deflections, the recalculated linewidths in Fig. \ref{highRes} are systematically {\it above} the standard measurement. This effect can be understood from the nonlinear pumping parameters $\Gamma_{\pm}$ that have been neglected here, but which do play a role at large displacements (see Fig. \ref{hnonlin}): the pumping factor decreases, which decreases the gain and thus the displacement $x_0$ at a fixed excitation force $F_0$. The resulting calculated linewidth in Eq. (\ref{eqDWres})  is then slightly overestimated. Taking into account the nonlinear term $(\Gamma_{+}-\Gamma_{-})/\Delta_{+}^0$ adjusted on Fig. \ref{hnonlin}, we obtain the dashed line in Fig. \ref{highRes}, in good agreement with the data.

As we increase the pumping strength $h_0$ and/or the displacement $x_0$, the device will eventually jump into the parametric oscillation regime. The highest value $h_0$ reachable for a given $x_0$ (i.e. a given driving force $F_0$) is what will finally limit the gain, and thus the quality of the parametric amplification scheme. 
In the linear theory, the gain diverges at $h_0 \rightarrow 1$. In practice, nonlinearities will trigger self-sustained oscillations before this limit, at a finite gain \cite{book2}.
In Ref. \cite{roukeshighfreq}, the thermal stability of the $Q$ factor (i.e. linewidth $\Delta \omega$) has been addressed. Clearly, {\it all} nonlinear effects have to be considered (anelasticity, geometrical terms, nonlinear pumping) for a comprehensive understanding of the actual threshold effect. Moreover, an additional experimental limitation has to be taken into account: as we approach $h(F_0,V_{dc} V_0) \rightarrow 1$, the stability of the experimentalist's pumping "knob" $V_{dc} V_0$ becomes critical. Noise $\delta V_{dc}$, $\delta V_{0}$ on the gate voltages can push the system into parametric self-sustained motion, without any way back.

Let us define $1-h = \epsilon $ with $0 < \epsilon(F_0,V_{dc} V_0) < 1$ a small (finite) number quantifying the distance to the actual threshold. Rewriting Eq. (\ref{hfit}) and injecting Eq. (\ref{Ypumph}), we obtain at second order in $x_0$:
\begin{eqnarray}
\epsilon & = & \left( 1 - h_0 \right) + h_0 \left( K+\frac{K'}{2} \right) \frac{ F_0 }{m_0 \Delta \omega_{00} \omega_{00} \, \epsilon } \nonumber \\
& & \!\!\!\!\!\!\!\!\!\! -h_0 
\left(  \frac{K'^2}{4}+ \frac{K \,K'}{2} + K^2 -  \frac{\beta_1}{\omega_{00}}- \frac{\beta_2}{\Delta \omega_{00}} + \frac{\Gamma_{+}-\Gamma_{-}}{\Delta_{+}^0} \right) \times \nonumber \\
& & \left( \frac{ F_0 }{m_0 \Delta \omega_{00} \omega_{00} \, \epsilon } \right)^2. \label{epsilon}
\end{eqnarray}
$\omega_{00}$ and $\Delta \omega_{00}$ are the linear characteristics of the device (obtained at $x_0 \rightarrow 0$), while $h_0$ is the corresponding pumping factor. It can be seen as the {\it imposed} pumping (fixed via $V_{dc} V_0$), while $h$ is the {\it truly obtained} pumping.
Note that rigorously speaking the expansion above is insufficient when $\epsilon$ becomes vanishingly small: higher-orders have necessarily to be taken into account. 
Keeping this point in mind, Eq. (\ref{epsilon}) can be recast in the form of a third order polynom:
\begin{equation}
\epsilon^3 - \left( 1 - h_0 \right) \, \epsilon^2 - h_0 a F_0 \, \epsilon  + h_0 b F_0^2 = 0 , \nonumber
\end{equation}
with $a$ and $b$ two numbers englobing all the nonlinearities: $a= ( K+K'/2)/(m_0 \Delta \omega_{00} \omega_{00})$, $b=[K'^2/4+K K'/2 +K^2 -\beta_1/\omega_{00}- \beta_2/\Delta \omega_{00} + (\Gamma_{+}-\Gamma_{-})/\Delta_{+}^0]/(m_0 \Delta \omega_{00} \omega_{00})^2$. Note that in the most general formalism thermal effects can be incorporated in these factors.  
Three roots exist, but only one has to be considered:
\begin{eqnarray}
\epsilon & = & \frac{1 - h_0}{3} + \nonumber \\
& & \frac{1}{3 \times 2^{1/3}} \Biggl[ \frac{2^{2/3}\,B}{ \left( A + \sqrt{-4 \, B^3+ A^2 } \,\right)^{1/3} } \Biggr. \nonumber \\
& &  \,\,\,\,\,\,\,\,\,\,\,\,\,\,\,\,\,\,\,\, \Biggl.  + \left( A + \sqrt{-4 \, B^3+ A^2 }\, \right)^{1/3} \Biggr],  \label{thres}
\end{eqnarray}
with:
\begin{eqnarray*}
A(F_0,h_0) & = & 2(1-h_0)^3 + 9 \, h_0 a (1-h_0)\, F_0 - 27 \, h_0 b \, F_0^2, \\
B(F_0,h_0) & = &  (1-h_0)^2 + 3 \, h_0 a \, F_0  .  
\end{eqnarray*}
In the $\left( h_0,F_0 \right)$ parameter plane, the device remains in the parametric amplification regime until $\epsilon$ becomes {\it complex-valued}. 

\begin{figure}[t!]
\includegraphics[height=6.5 cm]{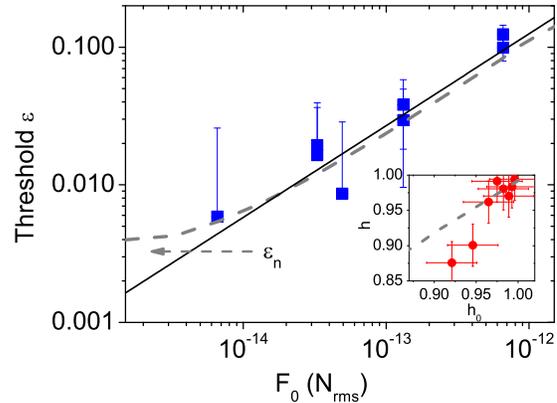}
\caption{\label{threshold} (Color online) 
Measured threshold $\epsilon=1-h$ as a function of the force $F_0$ for different pumping settings $h_0$. Inset: comparison between the obtained maximal pumping $h$ and the imposed pumping $h_0$.  The black line in the main graph is an $y=c'\,x^{2/3}$ dependence \cite{book2}, while the dashed lines arise from Eq. (\ref{thres}) fit on the data (see text). Noise on the gate voltages leads to $\epsilon_n$, main graph.}
\end{figure}

In Fig. \ref{threshold} we plot the measured $\epsilon$ at the threshold as a function of the force $F_0$ for various $h_0$. We obtain it from the fit linewidth $\Delta \omega$ just before self-sustained oscillations occur, either by keeping a given pumping $h_0$ and ramping the test force $F_0$, or by doing the reverse experiment. 
We know that the device switched to parametric oscillations from the {\it lineshape} of the resonance: the measured peak is not Lorentzian-looking anymore, and  Eqs. (\ref{Xpumph}) and (\ref{Ypumph}) cannot be used to fit the data. In the inset, Fig. \ref{threshold} we compare the imposed pumping $h_0$ and the obtained pumping $h(F_0,h_0)$ at the threshold.

The dashed lines in Fig. \ref{threshold} correspond to Eq. (\ref{thres}) adjusted numerically on the data, with an additional noise contribution $\epsilon_{n}$ arising from the finite stability of the gate voltages $\delta V_{dc}$, $\delta V_{0}$. We fit an average value  $\delta V/V$ of about $ 10^{-3}~$ which is reasonable regarding our experimental setup. Note that in practice, this finite value ensures that we do not use Eq. (\ref{epsilon}) outside of its applicability range. This parameter is also an ultimate limit for our implementation of the parametric amplification scheme. While the agreement between fit and data is fairly good, the obtained parameters $a$ and $b$ are rather badly defined; 
they only lie within the same order of magnitude as the ones calculated from the parameters $K$, $K'$, $\Gamma_{\pm}$ reported in the previous Section.
As a comparison, we show the power law $y=c'\,x^{2/3}$ derived for a single damping nonlinearity \cite{book2}, fit on the measurements. 
The agreement is again fairly good, which seems to indicate that the $x^{2/3}$ is a robust property obtained for {\it any} type of nonlinearities. However, we do not know how to link our nonlinear coefficients to the extracted fit parameter $c'$. 

\section{CONCLUSIONS}

In the present paper we report on experiments performed at low temperatures on nanoelectromechanical (NEMS) devices. The oscillators are 7$~$MHz cantilevered silicon structures of about 150$~$nm thickness and 3$~\mu$m length, actuated and detected through the magnetomotive scheme. A thin ($30~$nm) aluminum layer has been deposited to create electrical contacts. We implement  parametric amplification through a gate electrode capacitively coupled to the NEMS. 
A full modeling of the mechanical setup is presented, valid from the linear to the highly nonlinear range. \\
In the linear regime, both amplifying (gain $>1$) and squeezing (gain $<1$) have been demonstrated, as a function of the phase of the oscillating signals applied to the device. Gains up to about a 100 are obtained, for a structure which is {\it truly moving}: we explore the resonance from a tenth to about ten nanometer deflections (rms). At the gain maximum, forces of about a femtonewton are detected, for a structure moving about 2.6$~$\% of its thickness (peak-to-peak). Data are fit to theory, enabling an {\it in situ} measurement of the pumping strength. \\
Experiments are presented in the nonlinear regime, showing the reduction of the pumping strength for large displacements, up to about 50$~$nm. We demonstrate that the nonlinear behavior is renormalized by the gain.
We demonstrate the importance of the nonlinearity of the parametric pumping itself.
 Results are fit to theory, and we propose squeezing as a means to {\it reduce} the nonlinear behavior of NEMS. Nonlinear parametric amplification is used to study the {\it anelasticity} of the soft metallic coating, enabling the detection of the fine structure of the $Q$ factor: in our experiments we resolve the plastic limit of the aluminum film.\\
Finally, we studied the threshold between parametric amplification and oscillations. Our results are analyzed including {\it all} the nonlinear ingredients entering the modeling: geometrical nonlinearities, parametric pumping nonlinearities, materials nonlinearities.
Moreover, we introduce an unavoidable practical limit: the experimentalist's biases imprecision.
 We believe that our results constitute an important step in the understanding of the nonlinear behavior of parametrically amplified mechanical motion, and should help in particular  designing better devices.

\begin{acknowledgments}
We wish to thank T. Fournier, C. Lemonias, and B. Fernandez for their help in the fabrication of samples. We thank O. Exshaw and the electronics workshop for help in the measurement procedure.
The authors are grateful to J. Parpia for extremely valuable discussions.
We acknowledge the support from MICROKELVIN, the EU FRP7 low temperature infrastructure grant 228464.
\end{acknowledgments}

\begin{appendix}
\section{Defining the nonlinear coefficients from the dynamics equation}
\label{appendix}

In Section \ref{theory} we exposed the nonlinear equation describing parametric amplification at large deflections, Eq. (\ref{equation}). We assume that the 4th order pumping  "contamination" is negligible ($\alpha =0$) and that the spurious forces $F_{dc}$ and $F_{ac}$ are small enough to guarantee the integrity of the parametric amplification scheme (mathematically, same order as $x_0^2$ terms). Using the same method as the one presented in Ref. \cite{nonlinPRB}, we derive from a Mathematica code the nonlinear coefficients appearing in the harmonic response:
\begin{eqnarray*}
\beta_0 & = &  -\frac{1}{2} \left[ \tilde{k}_1 + \left( \frac{\omega}{\omega_0} \right)^2  \left( \frac{\Gamma_0}{m_0} - m_1 \right) \right] , 
\end{eqnarray*}
\begin{widetext}
\vspace*{-6 mm}
\begin{eqnarray*}
\beta_1 & = & + \frac{\omega_0}{2} \left[ \frac{3}{4} \tilde{k}_2 - \left( \frac{\omega}{\omega_0} \right)^2 \left( \frac{3}{4} m_2 - \frac{1}{4} \frac{\Gamma_1}{m_0} \right) + \left[ \frac{1}{2} m_1 \left( \frac{\omega}{\omega_0} \right)^2 - \tilde{k}_1 \right] \left[ \tilde{k}_1 + \left( \frac{\omega}{\omega_0} \right)^2 \left( \frac{\Gamma_0}{m_0} -m_1\right) \right] \right. \\
& &  \left. + \frac{(4\omega^2 - \omega_0^2)^2 \omega_0^2 \left[ \frac{1}{2} \left( \tilde{k}_1 - \frac{1}{2} \left( \frac{\omega}{\omega_0} \right)^2 \left[ 5 m_1 -4 \frac{\Gamma_0}{m_0} \right] \right) \left( \tilde{k}_1 - \left( \frac{\omega}{\omega_0} \right)^2 \left[ \frac{\Gamma_0}{m_0} +m_1 \right] \right) - \left( \frac{\omega \Lambda/m_0}{\omega_0^2}\right)^2 l_1^2  \right]}{(4\omega \Lambda/m_0)^2+(4\omega^2 -  \omega_0^2 )^2+2\frac{F_{dc}}{ 2 k_{foot} + k_0'} \left[ l_1 (4\omega \Lambda/m_0)^2   - (4\omega^2 -  \omega_0^2 )\, \omega_0^2 \left(2 \tilde{k}_1 - 4 m_1 \left(\frac{\omega}{\omega_0} \right)^2\right) \right]}   \right. \\
& & \left. + \frac{(4 \omega \Lambda /m_0)^2 \, \frac{3 l_1}{8} \left[- \tilde{k}_1+ \left( \frac{\omega}{\omega_0} \right)^2 \left( 2 m_1 - \frac{\Gamma_0}{m_0} \right)\right] }{(4\omega \Lambda/m_0)^2+(4\omega^2 -  \omega_0^2 )^2+2\frac{F_{dc}}{ 2 k_{foot} + k_0'} \left[ l_1 (4\omega \Lambda/m_0)^2   - (4\omega^2 -  \omega_0^2 )\, \omega_0^2 \left(2 \tilde{k}_1 - 4 m_1 \left(\frac{\omega}{\omega_0} \right)^2\right) \right]}   \right. \\
& & \left. + \frac{(4\omega^2 - \omega_0^2)^2 \omega_0^2 \, \frac{1}{2} \left[ \frac{\Delta k_1^2}{(2 k_{foot} + k_0')^2} -\frac{\Delta k_0\Delta k_1}{(2 k_{foot} + k_0')^2} \left(\tilde{k}_1 + \left( \frac{\omega}{\omega_0} \right)^2 \left[ \frac{\Gamma_0}{m_0} -m_1 \right] \right) \right] }{(4\omega \Lambda/m_0)^2+(4\omega^2 -  \omega_0^2 )^2+2\frac{F_{dc}}{ 2 k_{foot} + k_0'} \left[ l_1 (4\omega \Lambda/m_0)^2   - (4\omega^2 -  \omega_0^2 )\, \omega_0^2 \left(2 \tilde{k}_1 - 4 m_1 \left(\frac{\omega}{\omega_0} \right)^2\right) \right]}   \right], \\
\beta_1' & = & \frac{\omega_0}{2} \left[ -\frac{1}{2} \frac{\Delta k_2}{2 k_{foot} + k_0'} \sin (\phi') + \frac{\Delta k_1}{2 k_{foot} + k_0'} \times \right. \\
 & &\!\!\!\!\!\!\!\!\!\!\!\!\!\!\!\!\!\!\!\!\!\!\!\!\!\!\!\!\!\!\!\!\!\! \left. \frac{\left( -(4\omega^2 -  \omega_0^2 ) \sin (\phi') +(\frac{4\omega \Lambda}{m_0}) \cos (\phi') \right)\left( \tilde{k}_1 - \left( \frac{\omega}{\omega_0}\right)^2(\frac{\Gamma_0}{m_0}+m_1)\right)\omega_0^2 + l_1\left( \frac{2  \omega \Lambda }{m_0}\right) \left( +(4\omega^2 -  \omega_0^2 ) \cos (\phi') +(\frac{4\omega \Lambda}{m_0}) \sin (\phi') \right)}{(4\omega \Lambda/m_0)^2+(4\omega^2 -  \omega_0^2 )^2+2\frac{F_{dc}}{ 2 k_{foot} + k_0'} \left[ l_1 (4\omega \Lambda/m_0)^2   - (4\omega^2 -  \omega_0^2 )\, \omega_0^2 \left(2 \tilde{k}_1 - 4 m_1 \left(\frac{\omega}{\omega_0} \right)^2\right) \right]} \right], 
\end{eqnarray*}
\begin{eqnarray*}
\beta_2 & = & \left( \frac{\Lambda}{m_0}\right) \left[ \frac{1}{2} l_2 -l_1 \left( \tilde{k}_1 + \left( \frac{\omega}{\omega_0} \right)^2 \left[ \frac{\Gamma_0}{m_0} - m_1 \right] \right) \right. \\
& & \left. + \frac{(4\omega^2 - \omega_0^2)^2 \omega_0^2 \,l_1 \left( \frac{1}{2} \left[ \tilde{k}_1 -\left( \frac{\omega}{\omega_0} \right)^2  \left( \frac{\Gamma_0}{m_0} + m_1 \right)\right] +\tilde{k}_1 - \frac{1}{2} \left( \frac{\omega}{\omega_0} \right)^2  \left[ 5 m_1 - 4\frac{\Gamma_0}{m_0}  \right] \right)}{(4\omega \Lambda/m_0)^2+(4\omega^2 -  \omega_0^2 )^2+2\frac{F_{dc}}{ 2 k_{foot} + k_0'} \left[ l_1 (4\omega \Lambda/m_0)^2   - (4\omega^2 -  \omega_0^2 )\, \omega_0^2 \left(2 \tilde{k}_1 - 4 m_1 \left(\frac{\omega}{\omega_0} \right)^2\right) \right]} \right. \\
& & \left. + \frac{4 \omega_0^4 \left( \frac{1}{2} \left[\tilde{k}_1-\frac{1}{2} \left( \frac{\omega}{\omega_0} \right)^2 \left( 5 m_1 - 4 \frac{\Gamma_0}{m_0} \right) \right] \left[\tilde{k}_1-\left( \frac{\omega}{\omega_0} \right)^2 \left( \frac{\Gamma_0}{m_0} + m_1 \right) \right]-\left[ \frac{\omega \Lambda /m_0}{\omega_0^2 }\right]^2 l_1^2 \right) }{(4\omega \Lambda/m_0)^2+(4\omega^2 -  \omega_0^2 )^2+2\frac{F_{dc}}{ 2 k_{foot} + k_0'} \left[ l_1 (4\omega \Lambda/m_0)^2   - (4\omega^2 -  \omega_0^2 )\, \omega_0^2 \left(2 \tilde{k}_1 - 4 m_1 \left(\frac{\omega}{\omega_0} \right)^2\right) \right]} \right], \\
\beta_2' & = & \frac{\omega_0^2}{\omega} \left[  -\frac{1}{2} \frac{\Delta k_2}{2 k_{foot} + k_0'} \cos (\phi') + \frac{\Delta k_1}{2 k_{foot} + k_0'} \times \right. \\
 & &\!\!\!\!\!\!\!\!\!\!\!\!\!\!\!\!\!\!\!\!\!\!\!\!\!\!\!\!\!\!\!\!\!\! \left. \frac{\left( -(4\omega^2 -  \omega_0^2 ) \cos (\phi') -(\frac{4\omega \Lambda}{m_0}) \sin (\phi') \right)\left( \tilde{k}_1 - \left( \frac{\omega}{\omega_0}\right)^2(\frac{\Gamma_0}{m_0}+m_1)\right)\omega_0^2 + l_1\left( \frac{2  \omega \Lambda }{m_0}\right) \left( -(4\omega^2 -  \omega_0^2 ) \sin (\phi') +(\frac{4\omega \Lambda}{m_0}) \cos (\phi') \right)}{(4\omega \Lambda/m_0)^2+(4\omega^2 -  \omega_0^2 )^2+2\frac{F_{dc}}{ 2 k_{foot} + k_0'} \left[ l_1 (4\omega \Lambda/m_0)^2   - (4\omega^2 -  \omega_0^2 )\, \omega_0^2 \left(2 \tilde{k}_1 - 4 m_1 \left(\frac{\omega}{\omega_0} \right)^2\right) \right]} \right], \\ 
\delta \omega & = & \frac{\omega_0}{2} \left[ \frac{F_{dc}}{ 2 k_{foot} + k_0'} \left( 2 \tilde{k}_1 - m_1 \left(\frac{\omega}{\omega_0} \right)^2  \right) + \left(\frac{F_{dc}}{ 2 k_{foot} + k_0'}\frac{\Delta k_0\Delta k_1}{(2 k_{foot} + k_0')^2} - \frac{F_{ac}}{ 2 k_{foot} + k_0'} \frac{\Delta k_1}{2 k_{foot} + k_0'} \right) \times \right. \\
& & \left.  \frac{(4\omega^2 -  \omega_0^2 )\, \omega_0^2}{(4\omega \Lambda/m_0)^2+(4\omega^2 -  \omega_0^2 )^2+2\frac{F_{dc}}{ 2 k_{foot} + k_0'} \left[ l_1 (4\omega \Lambda/m_0)^2   - (4\omega^2 -  \omega_0^2 )\, \omega_0^2 \left(2 \tilde{k}_1 - 4 m_1 \left(\frac{\omega}{\omega_0} \right)^2\right) \right]} \right], \\
\delta (\Delta \omega ) & = & \frac{2 \Lambda}{m_0} \left( l_1 \frac{F_{dc}}{ 2 k_{foot} + k_0'} \right) , \\
\Delta_{+}^0 & = & \frac{\Delta k_0}{2 m_0}+\frac{\Delta k_1}{m_0} \frac{F_{dc} }{2 k_{foot} + k_0'} + \left( \frac{F_{ac}}{m_0} - \frac{F_{dc} }{m_0} \frac{\Delta k_0}{2 k_{foot} + k_0'}\right) \times \\
& & \frac{-(4\omega^2 -  \omega_0^2 )\, \omega_0^2 \left( \tilde{k}_1 + \left( \frac{\omega}{\omega_0} \right)^2 \left[2 \frac{\Gamma_0}{m_0} -\frac{5}{2} m_1\right] \right) +4 l_1 (\omega \Lambda/m_0)^2}{(4\omega \Lambda/m_0)^2+(4\omega^2 -  \omega_0^2 )^2+2\frac{F_{dc}}{ 2 k_{foot} + k_0'} \left[ l_1 (4\omega \Lambda/m_0)^2   - (4\omega^2 -  \omega_0^2 )\, \omega_0^2 \left(2 \tilde{k}_1 - 4 m_1 \left(\frac{\omega}{\omega_0} \right)^2\right) \right]},\\
\delta_{+}^0 & = & \frac{(\omega \Lambda/m_0) \, \omega_0^4 \left[ \frac{F_{ac}}{2 k_{foot} + k_0'} -  \frac{F_{dc} \Delta k_0}{(2 k_{foot} + k_0')^2} \right] \left[4 \left(\tilde{k}_1 + \left( \frac{\omega}{\omega_0} \right)^2 \left[ 2 \frac{\Gamma_0}{m_0} - \frac{5}{2} m_1 \right]\right) + l_1 (4\omega^2 -  \omega_0^2 )/\omega_0^2 \right]}{(4\omega \Lambda/m_0)^2+(4\omega^2 -  \omega_0^2 )^2+2\frac{F_{dc}}{ 2 k_{foot} + k_0'} \left[ l_1 (4\omega \Lambda/m_0)^2   - (4\omega^2 -  \omega_0^2 )\, \omega_0^2 \left(2 \tilde{k}_1 - 4 m_1 \left(\frac{\omega}{\omega_0} \right)^2\right) \right]}  , \\
\Gamma_{+} & = &  \frac{3}{8} \frac{\Delta k_2}{ m_0}+\frac{\Delta k_1}{2 m_0} \left(-\tilde{k}_1 + \left( \frac{\omega}{\omega_0} \right)^2 \left[ m_1- \frac{\Gamma_0}{m_0}\right] \right)  \\
& & \!\!\!\!\!\!\!\!\!\!\!\!\!\!\!\!\!\!\!\!\!\!\!\!\!\!\!\!\!\!\!\!\!\!\!\!\!\!\!\!\!\!\!\!\!\!\!\! + \frac{(4\omega^2 -  \omega_0^2 )\,\omega_0^2 \left[ - \frac{\Delta k_0}{m_0} \left(\frac{1}{2} \tilde{k}_1^2 +\tilde{k}_1\left( \frac{\omega}{\omega_0} \right)^2 \left[ \frac{3}{2}\frac{\Gamma_0}{m_0}-\frac{7}{4}  m_1 \right]  + \left( \frac{\omega}{\omega_0} \right)^4\left[ \left( \frac{\Gamma_0}{m_0} \right)^2 -\frac{9}{4} m_1 \frac{\Gamma_0}{m_0} +  \frac{5}{4} m_1^2\right] \right) + \frac{3}{4}\frac{\Delta k_1}{m_0} \left( \tilde{k}_1 + \left( \frac{\omega}{\omega_0} \right)^2 \left[\frac{\Gamma_0}{m_0} -2 m_1\right]  \right) \right] }{(4\omega \Lambda/m_0)^2+(4\omega^2 -  \omega_0^2 )^2+2\frac{F_{dc}}{ 2 k_{foot} + k_0'} \left[ l_1 (4\omega \Lambda/m_0)^2   - (4\omega^2 -  \omega_0^2 )\, \omega_0^2 \left(2 \tilde{k}_1 - 4 m_1 \left(\frac{\omega}{\omega_0} \right)^2\right) \right]} \\
& & \!\!\!\!\!\!\!\!\!\!\!\!\!\!\!\!\!\!\!\!\!\!\!\!\!\!\!\!\!\!\!\!\!\!\!\!\!\!\!\!\!\!\!\!\!\!\!\! + \frac{4 l_1 (\omega \Lambda /m_0)^2 \left[\frac{1}{2} \frac{\Delta k_0}{m_0} \left( \tilde{k}_1 - \left( \frac{\omega}{\omega_0} \right)^2 \left[ m1 - \frac{\Gamma_0}{m_0} \right]\right) -\frac{\Delta k_1}{m_0}  \right] }{(4\omega \Lambda/m_0)^2+(4\omega^2 -  \omega_0^2 )^2+2\frac{F_{dc}}{ 2 k_{foot} + k_0'} \left[ l_1 (4\omega \Lambda/m_0)^2   - (4\omega^2 -  \omega_0^2 )\, \omega_0^2 \left(2 \tilde{k}_1 - 4 m_1 \left(\frac{\omega}{\omega_0} \right)^2\right) \right]}, \\
\Gamma_{-} & = &\frac{1}{8} \frac{\Delta k_2}{ m_0} \\
& & + \frac{ \frac{\Delta k_1}{m_0} \left( \frac{1}{4}(4\omega^2 -  \omega_0^2 )\, \omega_0^2 \left[ \tilde{k}_1 - \left( \frac{\omega}{\omega_0} \right)^2  \left( \frac{\Gamma_0}{m_0} + m_1 \right)\right]-2 l_1 (\omega \Lambda/m_0)^2 \right)}{(4\omega \Lambda/m_0)^2+(4\omega^2 -  \omega_0^2 )^2+2\frac{F_{dc}}{ 2 k_{foot} + k_0'} \left[ l_1 (4\omega \Lambda/m_0)^2   - (4\omega^2 -  \omega_0^2 )\, \omega_0^2 \left(2 \tilde{k}_1 - 4 m_1 \left(\frac{\omega}{\omega_0} \right)^2\right) \right]} , 
\end{eqnarray*}
\begin{eqnarray*}
\gamma_{+} & = &\left( \frac{\omega \Lambda}{m_0}\right) \left[ \frac{ \frac{\Delta k_0}{m_0} \omega_0^2 \left[ 2 \tilde{k}_1^2 +\tilde{k}_1 \left( \frac{\omega}{\omega_0} \right)^2 \left( 6 \frac{\Gamma_0}{m_0} - 7 m_1\right) + \left( \frac{\omega}{\omega_0} \right)^4 \left( 4 \left[ \frac{\Gamma_0}{m_0}\right]^2 -9 m_1 \frac{\Gamma_0}{m_0} +5 m_1^2\right) \right]}{(4\omega \Lambda/m_0)^2+(4\omega^2 -  \omega_0^2 )^2+2\frac{F_{dc}}{ 2 k_{foot} + k_0'} \left[ l_1 (4\omega \Lambda/m_0)^2   - (4\omega^2 -  \omega_0^2 )\, \omega_0^2 \left(2 \tilde{k}_1 - 4 m_1 \left(\frac{\omega}{\omega_0} \right)^2\right) \right]} \right. \\
& & \left. + \frac{ \frac{1}{2} \frac{\Delta k_0}{m_0} (4\omega^2 -  \omega_0^2 ) \, l_1 \left[ \tilde{k}_1 -\left( \frac{\omega}{\omega_0} \right)^2 \left( m_1- \frac{\Gamma_0}{m_0} \right) \right] - \frac{\Delta k_1}{m_0} \omega_0^2 \left[  \tilde{k}_1 - \left( \frac{\omega}{\omega_0} \right)^2 \left( 4 m_1 - 5 \frac{\Gamma_0}{m_0} \right) \right]}{(4\omega \Lambda/m_0)^2+(4\omega^2 -  \omega_0^2 )^2+2\frac{F_{dc}}{ 2 k_{foot} + k_0'} \left[ l_1 (4\omega \Lambda/m_0)^2   - (4\omega^2 -  \omega_0^2 )\, \omega_0^2 \left(2 \tilde{k}_1 - 4 m_1 \left(\frac{\omega}{\omega_0} \right)^2\right) \right]} \right] , \\
\gamma_{-} & = &\left( \frac{\omega \Lambda}{m_0}\right) \frac{\frac{\Delta k_1}{ m_0}\, \omega_0^2 \left( \tilde{k}_1 +\frac{1}{2} l_1 (4\omega^2 -  \omega_0^2 )/\omega_0^2 - \left( \frac{\omega}{\omega_0} \right)^2 \left[ m_1 + \frac{\Gamma_0}{m_0} \right] \right) }{(4\omega \Lambda/m_0)^2+(4\omega^2 -  \omega_0^2 )^2+2\frac{F_{dc}}{ 2 k_{foot} + k_0'} \left[ l_1 (4\omega \Lambda/m_0)^2   - (4\omega^2 -  \omega_0^2 )\, \omega_0^2 \left(2 \tilde{k}_1 - 4 m_1 \left(\frac{\omega}{\omega_0} \right)^2\right) \right]} . 
\end{eqnarray*}
\end{widetext}
The above is the most generic result, valid for {\it any} quality factor $Q$. We recover the expressions of Ref. \cite{nonlinPRB} when the parametric pumping is turned off, as we should.

$k_{foot}$ and $\Lambda$ are $x_0$-dependent due to {\it anelastic} effects occurring in the materials. However, to a very good accuracy we can neglect in the above formulas the $k_{foot}$ variation, and consequently its repercussion in $\omega_0$ and $\tilde{k}_1,\tilde{k}_2$.
Even though the frequency tuning is quite strong, we always have  $k_0'/(2 k_{foot}) <\!<1$, which brings finally here $2 k_{foot} + k_0' \approx k_0$. 
In the high-$Q$ limit, these formulas can be further simplified, and in particular $\delta_{+}^0, \gamma_{+}, \gamma_{-}$ can be safely neglected. The explicit $\omega$-dependence appearing above disappears as well.

Moreover, we find experimentally that the {\it Duffing-like} nonlinear behavior of our devices is negligible (Fig. \ref{XvsF} insets), which means that $\tilde{k}_1,\tilde{k}_2, m_1,m_2, l_1, l_2, \Gamma_0,\Gamma_1$ terms can be disregarded; it leads to negligible $\beta_0, \beta_1, \beta_2$. The $\beta_1', \beta_2'$ terms are irrelevant on resonance $\omega = \omega_r$ since then $a^c_{1}  a^s_{1} = 0$. They can be neglected strictly if in addition to the weak intrinsic nonlinear behavior the normalized nonlinear pumping term $\Delta k_2 /(2 k_0)$ is sufficiently small. From the estimation of the capacitive expansion, Eq. (\ref{capa}) \cite{JAPtobe}, it is the case for our devices.

Similarly, the additional frequency shift and width terms $\delta \omega, \delta (\Delta \omega )$ can be safely ignored. We finally obtain at lowest order:
\begin{eqnarray*}
\Delta_{+}^0 & = & \frac{\Delta k_0}{2 m_0} ,\\
\Gamma_{+} & = &\frac{3}{8} \frac{\Delta k_2}{ m_0}, \\
\Gamma_{-} & = &\frac{1}{8} \frac{\Delta k_2}{ m_0} ,
\end{eqnarray*}
with the pumping factors $\Delta k_i$ ($i=0,2$) defined in Section \ref{theory}.

In practice for our nanomechanical systems, the {\it only} nonlinear effects that will affect in a relevant way the dynamics described here are the material-dependent terms, Eqs. (\ref{ko}), (\ref{lambda}) and the nonlinear pumping terms $\Gamma_{+},\Gamma_{-}$. As a result, our fit lineshapes always look Lorentzian; however in the most generic case, Duffing-like effects (combined in the $\beta_1$ term) have to be taken into account, and distort the resonance lines.

\end{appendix}



\begin{thebibliography}{blaaaaaaaaaaaaaaaaaaaaaaaaaaaaaaaaaaaaaaaaaaaa}


\bibitem{accel} J.D. Zook, W.R. Herb, C.J. Bassett, T. Stark, J.N. Schoess, and M.L. Wilson, {\it Sens. Actuators A} {\bf 83}, 270 (2000).
\bibitem{rugarnature} D. Rugar, R. Budakian, H.J. Mamin, and B.W. Chui, {\it Nature (London)} {\bf 430}, 329 (2004).
\bibitem{rugarnucl}  H.J. Mamin, M. Poggio, C.L. Deng, and D. Rugar, {\t Nat. Nanotechnol.} {\bf 2}, 301 (2007).
\bibitem{AFM} F.J. Giessibl, {\it Rev. Mod. Phys.} {\bf 75}, 949 (2003).
\bibitem{EFM} B.D. Terris, J.E. Stern, D. Rugar, and H.J. Mamin, {\it J. Vac. Sci. Technol. A} {\bf 8}, 374 (1989).
\bibitem{roukescleleand}  A.N. Cleland and M.L. Roukes, {\it Nature} {\bf 392}, 160 (1998). 
\bibitem{mohideen} U. Mohideen and A. Roy, {\it Phys. Rev. Lett.} {\bf 81}, 4549 (1998). 
\bibitem{mass1} K.L. Ekinci, X.M.H. Huang and M.L. Roukes, {\it Appl. Phys. Lett.} {\bf 84}, 4469–71 (2004); Y.T. Yang, C. Callegari, X.L. Feng, K.L. Ekinci, and M.L. Roukes, {\it Nano Lett.} {\bf 6} (4), pp 583 (2006).
\bibitem{mass2} B. Ilic, H.G. Craighead, S. Krylov, W. Senaratne, C. Ober and P. Neuzil, {\it J. Appl. Phys.} {\bf 95}, 3694–703 (2004). 
\bibitem{chemical} F.M. Battiston, J.P. Ramseyer, H.P. Lang, M.K. Baller, C. Gerber, J.K. Gimzewski, E. Meyer, and H.J. Guntherodt, {\it Sensors \& Actuators B-Chemical} {\bf  B77}, 122 (2001).
\bibitem {bio} B. Ilic, D. Czaplewski, M. Zalalutdinov, H.G. Craighead, P. Neuzil, C. Campagnolo, and C. Batt, {\it Journal of Vacuum Science \& Technology B} {\bf 19}, 2825 (2001).
\bibitem{schwab} M.D. LaHaye, O. Buu, B. Camarota, K.C. Schwab, {\it Science } {\bf 304}, 74 (2004). 
\bibitem{attonewton} H.J. Mamin and D. Rugar, {\it Appl. Phys. Lett.} {\bf 79}, 3358 (2001).
\bibitem{lehnert} J.D. Teufel, T. Donner, M.A. Castellanos-Beltran, J.W. Harlow and K.W. Lehnert, {\it Nature Nanotechnology} {\bf 4}, 820 (2009).
\bibitem{NIST} Michael J. Biercuk, Hermann Uys, Joe W. Britton, Aaron P. VanDevender, and John J. Bollinger, arXiv:1004.0780v4 [quant-ph] (2010).
\bibitem{clamp} Douglas M. Photiadis and John A. Judge, {\it Appl. Phys. Lett.} {\bf 85} (3), 482 (2004).
\bibitem{thermoel} Ron Lifshitz and M.L. Roukes, {\it Phys. Rev. B} {\bf 61} (8), 5600 (2000).
\bibitem{jeevakN} Scott S. Verbridge, Jeevak M. Parpia, Robert B. Reichenbach, Leon M. Bellan and H. G. Craighead, {\it J. Appl. Phys.} {\bf 99}, 124304 (2006).
\bibitem{highQ} Quirin P. Unterreithmeier, Thomas Faust, and J\"org P. Kotthaus, {\it Phys. Rev. Lett.} {\bf 105}, 027205 (2010).
\bibitem{tamayo} Javier Tamayo, {\it J. Appl. Phys.} {\bf 97}, 044903 (2005)
\bibitem{rugarPRL} D. Rugar, P. Gr\"utter, {\it Phys. Rev. Lett.} {\bf 67}, 699 (1991).
\bibitem{qcontrol} J. Kokavecz, Z. Horv\'ath, and A. M\'echl\'er, {\it Appl. Phys. Lett.} {\bf 85}, 3232 (2004).
\bibitem{jeevakvanderpohl} M. Zalalutdinov, A. Zehnder, A. Olkhovets, S. Turner, L. Sekaric, B. Ilic, D. Czaplewski,
J.M. Parpia, and H.G. Craighead, {\it Appl. Phys. Lett.} {\bf 79} (5), 695 (2001).
\bibitem{fiveparam} Kimberly L. Turner, Scott A. Miller, Peter G. Hartwell, Noel C. MacDonald, Steven H. Strogatz and Scott G. Adams, {\it Nature} {\bf 396}, 149 (1998).
\bibitem{nanowires} Min-Feng Yu and Gregory J. Wagner, {\it Nanotech.} {\bf 3}, 325 (2003); Min-Feng Yu, Gregory J. Wagner, Rodney S. Ruoff, and Mark J. Dyer, {\it Phys. Rev. B}  {\bf 66}, 073406 (2002).
\bibitem{stmparam} M. Moreno-Moreno, A. Raman, J. Gomez-Herrero, R. Reifenberger, {\it Appl. Phys. Lett.} {\bf 88}, 193108 (2006).
\bibitem{swing} S. M. Curry, Am. J. Phys. 44, p 924 (1976); J.A. Burns, {\it Am. J. Phys.} {\bf 38}, 920 (1970).
\bibitem{water} M. Faraday, {\it Philos. Trans. R. Soc. London} {\bf 121}, 319 (1831).
\bibitem{elec} D.P. Howson, and R.B. Smith, {\it Parametric Amplifiers} (McGraw-Hill, New York, 1970).
\bibitem{josephson} R. Movshovich, B. Yurke, P.G. Kaminsky, A.D. Smith, A.H. Silver, R.W. Simon, M.V. Schneider, {\it Phys. Rev. Lett.} {\bf 65}, 1419 (1990).
\bibitem{opo} Gian-Luca Oppo, Massimo Brambilla and Luigi A. Lugiato, {\it Phys. Rev. A} {\bf 49}, 2028 (1994).
\bibitem{squeezedlight} L.A. Wu, H.J. Kimble, J.L. Hall, and H. Wu, {\it Phys. Rev. Lett.} {\bf 57}, 2520 (1986).
\bibitem{book} A.H. Nayfeh and D.T. Mook, {\it Nonlinear Oscillations} (Wiley, New York, 1979).
\bibitem{book2} Heinz Georg Schuster Ed., {\it Reviews of Nonlinear Dynamics and Complexity}, Vol. 1, Chapter I by R. Lifshitz and M. C. Cross, Wiley-VCH (2008).
\bibitem{pumphighQ} I. Mahboob and H. Yamaguchi, {\it Appl. Phys. Lett.} {\bf 92}, 253109 (2008).
\bibitem{macro} Jeffrey F. Rhoads, Nicholas J. Miller, Steven W. Shaw, Brian F. Feeny, 
{\it Proceedings of the ASME 2007 International Design Engineering Technical Conferences \& Computers and Information in Engineering Conference}, 35426 (2007).
\bibitem{gaasstress} A. D\^ana, F. Ho, Y. Yamamoto, {\it Appl. Phys. Lett.} {\bf 72}(10), 1152 (1998).
\bibitem{torsionalparpia} Dustin W. Carr, Stephane Evoy, Lidija Sekaric, H.G. Craighead, and J.M. Parpia, {\it Appl. Phys. Lett.} {\bf 77} (10), 1545 (2000).
\bibitem {parpiadisk} M. Zalalutdinov, A. Olkhovets, A. Zehnder, B. Ilic, D. Czaplewski, H.G. Craighead, and J.M. Parpia, {\it Appl. Phys. Lett.} {\bf 78} (20), 3142 (2001).
\bibitem{roukeshighfreq} R.B. Karabalin, X.L. Feng, and M.L. Roukes, {\it Nano Letters} {\bf 9} (9), 3116 (2009).
\bibitem{schwabqubit} Junho Suh, Matthew D. LaHaye, Pierre M. Echternach, Keith C. Schwab, and Michael L. Roukes, {\it Nano Letters} {\bf 10}, 3990 (2010).
\bibitem{highgain} R.B. Karabalin, S.C. Masmanidis, and M.L. Roukes, {\it Appl. Phys. Lett.} {\bf 97}, 183101 (2010).
\bibitem{chevrier} T. Ouisse, M. Stark, F. Rodrigues-Martins, B. Bercu, S. Huant, and J. Chevrier, {\it Phys. Rev. B} {\bf 71}, 205404 (2005).
\bibitem{cleland} A.N. Cleland, {\it New Journal of Physics} {\bf 7}, 235 (2005).
\bibitem{book3} A.N. Cleland, {\it Foundations of Nanomechanics}, Springer 2003.
\bibitem{nlinpaper0} Wenhua Zhang, Rajashree Baskaran, Kimberly L. Turner, {\it Sensors and Actuators A} {\bf 102}, 139 (2002).
\bibitem{nlinpaper1} Wenhua Zhang, Rajashree Baskaran, and Kimberly Turner, {\it Appl. Phys. Lett.} {\bf 82}, (1), 130 (2003).
\bibitem{nlinpaper2} Jeffrey F Rhoads, Steven W Shaw, and Kimberly L Turner, {\it J. Micromech. Microeng.} {\bf 16}, 890 (2006).
\bibitem{nlinpaper3} Jeffrey F. Rhoads, Steven W. Shaw, Kimberly L. Turner, Jeff Moehlis, Barry E. DeMartini, Wenhua Zhang, {\it Journal of Sound and Vibration} {\bf 296}, 797 (2006).
\bibitem{nlinpaper4} Barry E. DeMartini, Jeffrey F. Rhoads, Kimberly L. Turner, Steven W. Shaw, and Jeff Moehlis, {\it J. of Microelectromech. Syst.} {\bf 16} (2),  310 (2007).
\bibitem{qfs} E. Collin, T. Moutonet, J.-S. Heron, O. Bourgeois, Yu.M. Bunkov, H. Godfrin, {\it J. of Low Temp. Phys.} {\bf 162} (5/6),  653 (2011).
\bibitem{JLTP_VIW} Eddy Collin, Laure Filleau, Thierry Fournier, Yuriy M. Bunkov and Henri Godfrin, {\it J. of  Low Temp. Phys.} {\bf 150}, 739 (2008); Erratum {\bf 157},  566 (2009).
\bibitem{qfs2009} E. Collin, J. Kofler, J.-S. Heron, O. Bourgeois, Yu.M. Bunkov, H. Godfrin, {\it J. of Low Temp. Phys.} {\bf 158} (3), 678 (2010). 
\bibitem{nonlinPRB} E. Collin, Yu.M. Bunkov, H. Godfrin, {\it Phys. Rev. B}  {\bf 82}, 235416 (2010).
\bibitem{coatings} E. Collin, J. Kofler, S. Lakhloufi, S. Pairis, Yu.M. Bunkov, and H. Godfrin, {\it J. Appl. Phys.} {\bf 107} (11), 114905 (2010).
\bibitem{note} Using the Rayleigh method, we estimate the correction to the detected signal as a function of the deflection amplitude $x_0$. We compute the $\cos \theta$ between the field direction and the distorted beam. The linear regime Rayleigh expression is $f(y,t) = x(t) \, [3/2 (y/l')^2-1/2 (y/l')^3]$. Thus $\cos  \theta= 1 - 9/8 (x/l')^2$ at lowest order. For a displacement of 50$~$nm, the error is well below $1~$\%. Moreover, the shortening of the cantilever feet abscissa can be estimated with $y = -  3/5 \, l' \, (x/l')^2$. The exact mode shape will only modify very little these numbers. But note that to be perfectly rigorous, intrinsic cantilever nonlinearities will add cubic terms to these expressions \cite{nonlinPRB}. In our case, this is all negligible.
\bibitem{tuningRoukes} I. Kozinsky, H.W.Ch. Postma, I. Bargatin, and M.L. Roukes, {\it Applied Physics Letters} {\bf 88}, 253101 (2006).
\bibitem{JAPtobe} Submitted to {\it Journal of Applied Physics}.
\bibitem{landau} L.D. Landau \& E.M. Lifshitz, {\it Mechanics}, Third Ed. Elsevier Science Ltd. (1976).
\end{thebibliography}
\end{document}